\documentclass[aps,prb,reprint]{revtex4-2}

\usepackage{bm}
\usepackage{graphicx}

\begin{document}


\title{System-bath model for quantum chemistry}

\author{Dmitry S. Golubev$^1$}

\author{Reza G. Shirazi$^1$}
	
\author{Vladimir V. Rybkin$^1$}

\author{Benedikt M. Schoenauer$^1$}

\author{Peter Schmitteckert$^1$}
	
\author{Michael Marthaler$^1$}

\affiliation{$^1$HQS Quantum Simulations GmbH, Rintheimer Str. 23, 76131 Karlsruhe, Germany}

\begin{abstract}
We propose an approximate mapping of a molecular Hamiltonian to a Hamiltonian of qubits, which allows 
for high accuracy quantum chemistry calculations of vertical excitation energies of some molecules. The mapping
is based on separating of a very small active space of only two orbitals and on modeling the 
electronic excitations in the remaining orbitals by a set of qubits or, equivalently, by a set of oscillators.
This approach is inspired by the Random Phase Approximation (RPA),
in which the excitations of electron gas are described by bosonic degrees of freedom. 
As a result, the Hamiltonian of the molecule is reduced to that of a 
system-bath model. The "system" part of the Hamiltonian describes the two molecular orbitals --
the highest occupied molecular orbital (HOMO) and the lowest unoccupied molecular orbital (LUMO) -- which are populated by two electrons.
Two qubits are sufficient to encode the Hamiltonian of such a system. The "bath"
consists of oscillators or, equivalently, of two level systems with each of them corresponding to an electron excitation
from a doubly occupied orbital below the Fermi level to an empty orbital above the Fermi level. We hope that this mapping can inspire
new approaches and algorithms aimed at calculating excitation energies of molecules on near term quantum computers. 
\end{abstract}
	
\maketitle

\section{Introduction}

We present a way of transforming the molecular electronic-structure Hamiltonian to a form closely resembling 
the Hamiltonian of the well-known spin--boson model. The transformed Hamiltonian provides a convenient
starting point for quantum computing simulations. 
Early proof-of-principle quantum simulations of molecular electronic structure have already been demonstrated on gate-based hardware,
including Hartree--Fock simulations for hydrogen chains~\cite{Google} and the singlet--triplet gap of methylene diradical~\cite{methylene_QC}.
Theoretically proposed advanced quantum computing algorithms for chemistry are
usually based on Jordan–Wigner transformation or its extensions \cite{Jiang,McArdle,Li}, 
which exactly express fermionic operators in terms of
qubit operators.

\begin{figure}[t]
\begin{center}
\begin{tabular}{ccc}
singlet molecules\hspace{1cm} & \hspace{0.5cm} & triplet molecules\hspace{1cm} \\
                  & \hspace{0.5cm} &                   \\
\includegraphics[width=0.4\columnwidth]{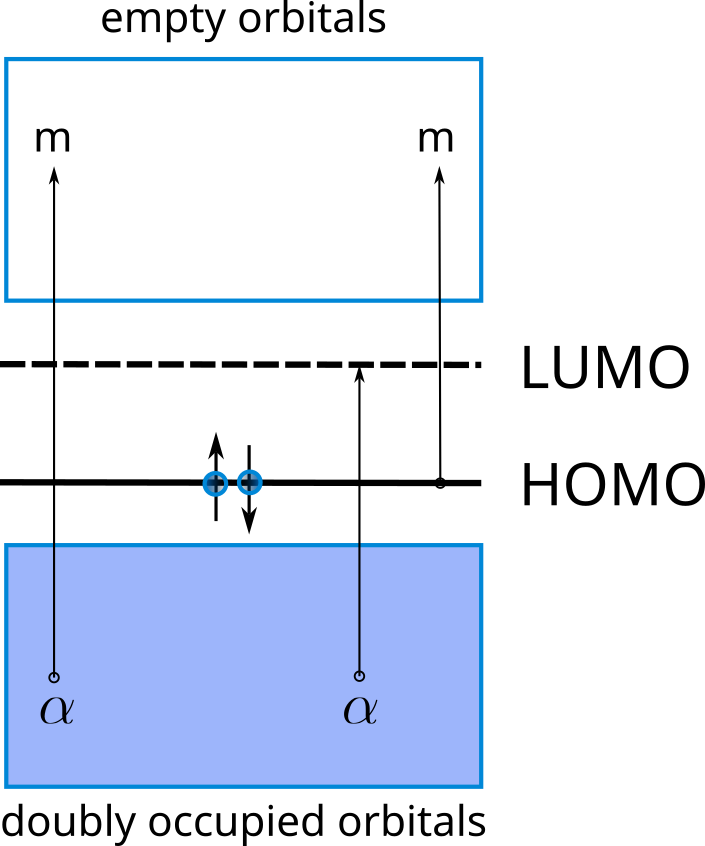} & \hspace{0.5cm} & \includegraphics[width=0.4\columnwidth]{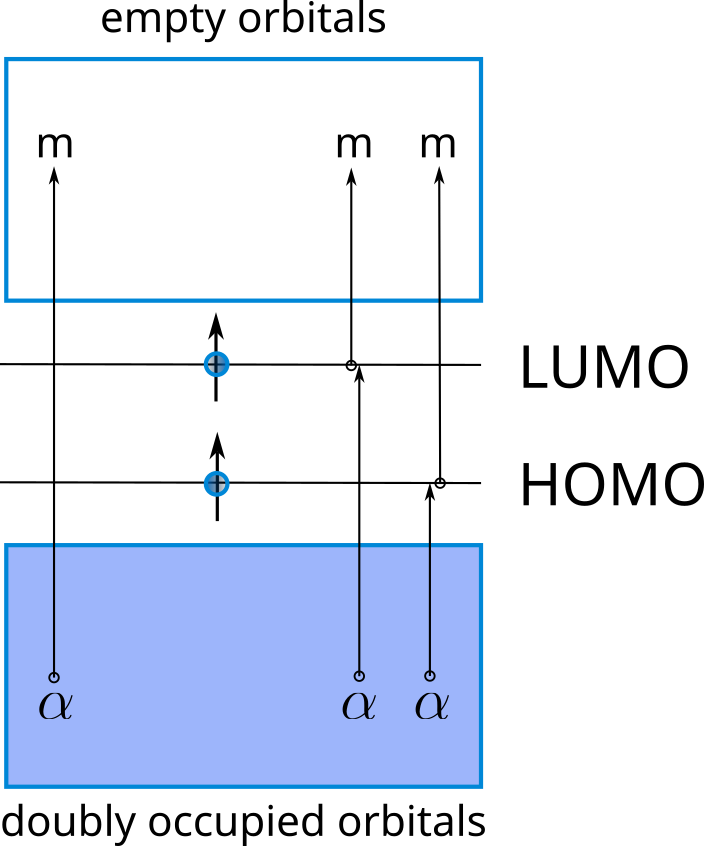}
\end{tabular}
\end{center}
\caption{Possible transitions between the environment orbitals (arrows in the left side of both figures) and between the environment orbitals
and the HOMO/LUMO orbitals (arrows in the right side of both figures). Each of the transitions corresponds to an oscillator  
in the bath. Left figure refers to the molecules with the singlet ground state of the isolated HOMO/LUMO orbitals, and the 
right figure --- to the molecules with the triplet ground state of the isolated HOMO/LUMO orbitals.}
\label{Fig_exc}
\end{figure}

In contrast to this earlier work, we focus here on constructing an approximate Hamiltonian form that is substantially more amenable to near-term quantum simulation. 
Instead of mapping a large fermionic active space directly onto a quantum computer, we show that a partitioning of the orbital space allows 
one to rewrite the problem as an effective ``system + environment'' Hamiltonian,
where the environment is represented by bosonic modes describing orbital transitions.
Concretely, we arrive at the decomposition
\begin{equation}
H \;=\; H_{\mathrm{sys}}  \;+\; H_{\mathrm{env}} \;+\; \tilde V.
\end{equation}
Here, $H_{\mathrm{sys}}$ acts on a small, fixed-dimensional electronic subspace (in this work, the two-electron HOMO/LUMO subspace),
$H_{\mathrm{env}}$ describes the electron excitations in the remaining orbitals in terms of effective bosonic modes (see Fig. \ref{Fig_exc}),
and $\tilde V$ couples the two.

\begin{table}[b]
\begin{tabular}{|c|c|c|c|c|}
\hline
molecule        & static     & 62 qubits  & 126 qubits &  MR-AQCC    \\
                & app.       &            &            &           \\
\hline
cyclopentadiene & 3.408      & 3.331      & 3.314      & 3.326      \\
\hline
pyrrole         & 4.588      & 4.485      & 4.463      & 4.487      \\
\hline
thiophene       & 4.066      & 3.969      & 3.926      & 3.935      \\
\hline
\end{tabular}
\caption{Vertical excitation energies of the lowest triplet states (eV) for
three molecules calculated with the
static approximation of Sec. \ref{Sec_static}, by diagonalizing the model Hamiltonian with 62 and 126 qubits in the environment, 
and with the accurate multireference averaged quadratic coupled cluster (MR-AQCC) method.
More details are given in Sec. \ref{results}. }
\label{table222}
\end{table}

The key point is that the interaction can be cast into a compact, spin--boson-like form,
\begin{eqnarray}
\tilde V
&=&
\sum_{m\alpha}\!\left(b_{m\alpha}^\dagger+b_{m\alpha}\right)
\left[
g_{12}^{m\alpha}\,\hat O^{x} \right. \nonumber \\
& &+
\left. \frac{g_{11}^{m\alpha}-g_{22}^{m\alpha}}{2}\,\hat O^{z}
\right]
\;+\;
V_{\mathrm{ct}}.
\end{eqnarray}
Here $b_{m\alpha}$ is the ladder operator of the environment oscillator approximately corresponding
to an electron excitation from a doubly occupied orbital $\alpha$ to an empty orbital $m$ (see Fig. \ref{Fig_exc}),
$g_{12}^{m\alpha}$ and $g_{11}^{m\alpha},g_{22}^{m\alpha}$ are the transverse and the longitudinal coupling constants, and
the operators $\hat O^{x}$ and $\hat O^{z}$ are the system operators responsible for the generalized transverse and longitudinal couplings respectively.
They are analogous in spirit to the familiar Pauli operators $\sigma_x$ and $\sigma_z$ of the standard spin--boson model,
but here they act on a four-dimensional electronic subspace (they are $4\times 4$ matrices) associated with the two-orbital/two-electron HOMO/LUMO manifold. 
The term $V_{\mathrm{ct}}$ contains the counter terms and the associated Lamb-shift contributions required for a well-defined continuum limit of the environment
(and, in practice, for stable convergence when the number of modes is increased), mirroring a familiar feature of spin--boson physics.

Despite its compact form, the resulting model can be quantitatively accurate.
As demonstrated in Table \ref{table222}, even with a very simplistic truncation of the environmental modes, namely with reducing their number to 62 or 126
and absorbing the effect of all remaining modes into the renormalized parameters of the system,  
the approach reproduces reference excitation energies with excellent accuracy for three representative closed shell molecules.
This highlights that the system--bath representation can retain chemically relevant accuracy while providing a Hamiltonian structure that is significantly more amenable to near-term quantum simulations.
System--bath (spin--boson) Hamiltonians have a long history as effective models for dissipation, decoherence, and renormalization effects in quantum dynamics, with comprehensive treatments given in Refs.~\cite{Leggett,Weiss}. 
On the quantum-chemistry side, representing collective electron--hole excitations by effective bosonic modes is closely connected to established many-body ideas 
such as the random-phase approximation and related treatments of electronic response and correlation~\cite{Bohm1,Bohm2,Bohm3,Furche}, 
and provides a complementary route of incorporating dynamical correlation beyond minimal active-space pictures~\cite{Meyer,Szabo,Scott1,Scott,RPA}.

This reformulation directly addresses two major obstacles that arise when attempting to run quantum chemistry on quantum hardware.
First, the molecular electronic Hamiltonian is dominated by fermionic two-body interactions with four orbital indices,
leading after fermion-to-qubit mappings to very large numbers of Pauli terms and deep circuits.
Second, small active-space models often neglect (or only partially capture) dynamical correlation effects due to the remaining orbitals,
even though such effects can be essential for quantitative description of excitation energies and singlet--triplet splittings.

\begin{figure}[t]
\begin{center}
\includegraphics[width=0.95\columnwidth]{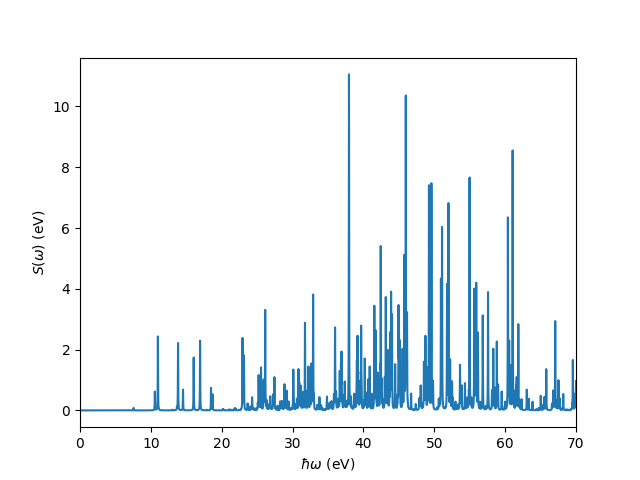} 
\end{center}
\caption{Discrete transverse coupling spectrum associated with the system--bath mapping for thiophene molecule.
We plot the quantity
$S(\omega)=\sum_{m\alpha}\big(g_{12}^{m\alpha}\big)^2\,\delta\!\left(\omega-\Omega_{m\alpha}\right)$,
i.e., a spectral-density-like representation of the mode frequencies $\Omega_{m\alpha}$ and the transverse couplings $g_{12}^{m\alpha}$.
$\delta$-functions are replaced by Lorentzian peaks with the half-width 0.02 eV.
$S(\omega)$ is directly analogous to the spectral density in the spin--boson models, which motivates importing
spin--boson discretization/encoding techniques~\cite{Miessen2021,Burger2022,Tudorovskaya2024,Walters2024,LasHeras2015} into the present setting.}
\label{Fig_spectral}
\end{figure}

At the same time, a large body of work in quantum algorithms and quantum simulation focuses on efficient gate-based treatments of
spin--boson-type models and closely related system--bath Hamiltonians.
Our approach leverages this situation: it maps a relevant quantum-chemistry target (low-lying excitation energies of molecules) onto a Hamiltonian family
that is substantially closer to those models that are already intensively studied in quantum computing community.
In particular, many of these studies develop techniques to discretize, encode, or otherwise optimize representations of bosonic environments
and their spectral densities in gate-based circuits~\cite{Miessen2021,Burger2022,Tudorovskaya2024,Walters2024,LasHeras2015, HQS1, HQS2, HQS3}.
This opens the door to importing such techniques into the present quantum-chemistry mapping, where the distribution of mode frequencies and
couplings (illustrated in Fig.~\ref{Fig_spectral}) plays an analogous role to a spin--boson spectral density.

Our model is based on separating the molecular orbital space into a minimal ``system'' subspace and a large ``environment'' subspace.
In the applications considered here, the system is formed by the HOMO and LUMO populated by two electrons,
while all remaining orbitals constitute the environment.
Excitations within the environment are represented by bosonic modes, and their coupling to the HOMO/LUMO subsystem is derived from the underlying electronic interaction,
resulting in the effective coupling constants $g_{ij}^{m\alpha}$.
This mapping is inspired by random-phase-approximation  and yields a controlled way of incorporating dynamical screening and correlation effects beyond the static active-space picture.
Figure \ref{Fig_exc} illustrates the relevant classes of transitions between system and environment orbitals and motivates the chosen partitioning.

\section{Hamiltonian in simple terms}
\label{sec:hamiltonian_bridge}

The system-bath Hamiltonian resulting from the mapping of a molecular Hamiltonian is
rather complicated, it will be presented in its complete form in Sec. \ref{qubit_env}. 
In this section we briefly describe its structure
and its derivation in simple terms. 

The starting point of our construction is the standard second-quantized molecular Hamiltonian,
which contains one-electron terms and two-electron repulsion integrals with four orbital indices.
Direct fermion-to-qubit mappings for such a Hamiltonian typically generate a large number of Pauli terms and deep circuits.

To arrive at a more hardware-friendly structure, we partition the orbital space into a minimal ``system'' subspace and an ``environment'' subspace.
In this work the system consists of the HOMO and LUMO orbitals populated by two electrons (indices $1,2$),
while all remaining orbitals form the environment.
This yields the exact decomposition
\begin{equation}
H = H_{12} + H_{\rm env} + V,
\end{equation}
with $H_{12}$ being the HOMO/LUMO Hamiltonian [Eq.~(\ref{H0})], $H_{\rm env}$ the environment Hamiltonian [Eq.~(\ref{Henv})],
and $V$ the interaction [Eqs.~(\ref{Vint})--(\ref{V3})].

Averaging the interaction over the environment occupations (Hartree--Fock reference for the environment) produces an effective system Hamiltonian
\begin{equation}
H_{\rm sys} = H_{12} + \langle V\rangle 
\end{equation}
[Eq.~(\ref{Hsys})], and defines the renormalized interaction $\tilde V$ via
\begin{equation}
H = H_{\rm sys} + H_{\rm env} + \tilde V
\end{equation}
[Eq.~(\ref{H_full})] with $\tilde V= V - \langle V\rangle$ given in Eq.~(\ref{V_renorm}).

In Sec.~\ref{model} and in the Appendices we approximate, inspired by RPA, the environment excitations by a set of harmonic modes,
diagonalize the coupled-oscillator problem, and express the environment in terms of normal modes with ladder operators $b_{m\alpha}$ and
frequencies $\Omega_{m\alpha}$ [Eq.~(\ref{Henv_diag})]. The system part $H_{\rm sys}$ is mapped to a two-qubit Hamiltonian
acting on the four-dimensional HOMO/LUMO two-electron manifold [Eq.~(\ref{H_sys_qubits})].
Finally, the interaction $\tilde V$ is expressed in the normal-mode basis and rewritten in a compact ``spin--boson-like'' form [Eq.~(\ref{V_int_normal})],
which is provided already in the Introduction,
\begin{equation}
\tilde V
=
\sum_{m\alpha}\left(b_{m\alpha}^\dagger+b_{m\alpha}\right)
\left[
g_{12}^{m\alpha}\,\hat O^{x}
+
\frac{g_{11}^{m\alpha}-g_{22}^{m\alpha}}{2}\,\hat O^{z}
\right]
+ V_{\rm ct}.
\end{equation}
Here $\hat O^{x}$ and $\hat O^{z}$ are system operators acting in the $4\times 4$ HOMO/LUMO subspace (generalized transverse/longitudinal couplings),
and $V_{\rm ct}$ collects the counter-term and associated Lamb-shift contributions (the system-only renormalization terms) required for correct normal ordering
and stable convergence in the static-screening/continuum-bath limit (see Sec.~\ref{counter} and the discussion below Eq.~(\ref{V_int_normal})).

\section{Model}

\subsection{Hamiltonian of a molecule}
\label{Ham}

The non-relativistic Hamiltonian of a molecule has the following general form
\begin{eqnarray}
&& H = \sum_\sigma\int d^3{\bm r}\; \Psi_\sigma^\dagger({\bm r})\left[ -\frac{\hbar^2\nabla^2}{2m} + U_{\rm nucl}({\bm r}) \right]\Psi_\sigma({\bm r})
\nonumber\\ &&
+\, \frac{e^2}{2} \sum_{\sigma\sigma'}\int d^3{\bm r} d^3{\bm r}' \; \frac{\Psi_\sigma^\dagger({\bm r})\Psi_{\sigma'}^\dagger({\bm r}')\Psi_{\sigma'}({\bm r}') \Psi_\sigma({\bm r})}{|{\bm r}-{\bm r}'|}.
\label{HH}
\end{eqnarray}
Here $\Psi_\sigma({\bm r}),\Psi_\sigma^\dagger({\bm r})$ are the annihilation and creation operators 
of an electron at point ${\bm r}$ with the spin projection on $z$-axis $\sigma$, $e$ is the electron charge, $\hbar$ is the Planck's constant
and $U_{\rm nucl}({\bm r})$ is the Coulomb potential of the atomic nuclei. 
Introducing the orthogonal basis of real valued wave functions of the molecular orbitals $\psi_\sigma({\bm r})$, we write the electronic operators in the form
\begin{eqnarray}
\Psi_\sigma({\bm r}) = \sum_{p,\sigma} \psi_\sigma({\bm r}) c_{p,\sigma}, \;\;\; \Psi_\sigma^\dagger({\bm r}) = \sum_{p,\sigma} \psi_\sigma({\bm r}) c^\dagger_{p,\sigma},
\end{eqnarray}
where $c_{p\sigma},c_{p\sigma}^\dagger$ are the annihilation and creation operators for the orbitals.
In terms of these operators, the Hamiltonian (\ref{H}) is expressed as
\begin{eqnarray}
H = \sum_\sigma\sum_{pq} t_{pq} c^\dagger_{p\sigma} c_{q\sigma} 
+ \frac{1}{2} \sum_{\sigma\sigma'} \sum_{pqrs} h_{pqrs} c^\dagger_{p\sigma} c^\dagger_{r\sigma'} c_{s\sigma'} c_{q\sigma}, 
\label{H}
\end{eqnarray}
where the single particle amplitudes $t_{pq}$ and the "chemical" electron repulsion integrals $h_{pqrs}$ read 
\begin{eqnarray}
&& t_{pq} = \int d^3{\bm r}\; \psi_p({\bm r})\left[ -\frac{\hbar^2\nabla^2}{2m} + U_{\rm nucl}({\bm r}) \right]\psi_q({\bm r}),
\label{t_pq}\\
&& h_{pqrs} = e^2\int d^3{\bm r} d^3{\bm r}' \frac{\psi_p({\bm r})\psi_q({\bm r}) \psi_r({\bm r}')\psi_s({\bm r}') }{|{\bm r}-{\bm r}'|}.
\label{eri}
\end{eqnarray}

We separate the HOMO and LUMO orbitals, treating them as "system", 
assign the index 1 to the HOMO orbital, the index 2 to the LUMO orbital
and assume that only two electrons populate them. 
All remaining molecular orbitals form the environment,
and we denote the total number of the environment orbitals for a given basis set as $N_{\rm env}$.
In accordance with that, we split the Hamiltonian into three parts,
\begin{eqnarray}
H = H_{12} + H_{\rm env} + V,
\end{eqnarray} 
where
\begin{eqnarray}
&& H_{12} = \sum_\sigma\sum_{p_1p_2=1}^{2} t_{p_1p_2} c^\dagger_{p_1\sigma} c_{p_2\sigma} 
\nonumber\\ &&
+ \frac{1}{2} \sum_{\sigma\sigma'} \sum_{p_1p_2p_3p_4=1}^{2} h_{p_1p_2p_3p_4} c^\dagger_{p_1\sigma} c^\dagger_{p_3\sigma'} c_{p_4\sigma'} c_{p_2\sigma}, 
\label{H0}
\end{eqnarray}
is the Hamiltonian of the HOMO/LUMO orbitals,  
\begin{eqnarray}
&& H_{\rm env} = \sum_\sigma\sum_{k_1k_2=1}^{N_{\rm env}} t_{k_1k_2} c^\dagger_{k_1\sigma} c_{k_2\sigma} 
\nonumber\\ &&
+ \frac{1}{2} \sum_{\sigma\sigma'} \sum_{k_1k_2k_3k_4=1}^{N_{\rm env}} h_{k_1k_2k_3k_4} c^\dagger_{k_1\sigma} c^\dagger_{k_3\sigma'} c_{k_4\sigma'} c_{k_2\sigma}, 
\label{Henv}
\end{eqnarray}
is the Hamiltonian of the environment orbitals, and $V$ is the interaction term, which 
can be further split into several parts as \cite{RPA}, 
\begin{eqnarray}
V = V_t + V_{\rm C} + V_{\rm ex} + V_{\rm pair} + V_1 + V_3.
\label{Vint}
\end{eqnarray}
Here
\begin{eqnarray}
V_t = \sum_\sigma\sum_{p=1}^{2} \sum_{k=3}^{N_{\rm env}} t_{pk} ( c^\dagger_{p\sigma} c_{k\sigma} + c^\dagger_{k\sigma} c_{p\sigma} )
\label{V_t}
\end{eqnarray}
describes the hopping between the HOMO/LUMO orbitals and the environment,
\begin{eqnarray}
V_{\rm C} = \sum_{\sigma,\sigma'}\sum_{p_1p_2=1}^{2} \sum_{k_1k_2=1}^{N_{\rm env}} h_{p_1p_2 k_1k_2}  c^\dagger_{p_1\sigma} c_{p_2\sigma}  c^\dagger_{k_1\sigma'} c_{k_2\sigma'}  
\label{V_C}
\end{eqnarray}
is the Coulomb interaction between the HOMO/LUMO orbitals and the environment,
\begin{eqnarray}
&& V_{\rm ex} = -  \frac{1}{2}\sum_{\sigma\sigma'} \sum_{p_1p_2=1}^{2} \sum_{k_1k_2=1}^{N_{\rm env}} 
h_{p_1 k_1 p_2 k_2} 
\nonumber\\ &&\times\,
\big( c^\dagger_{p_1\sigma} c_{p_2\sigma'} c^\dagger_{k_2\sigma'} c_{k_1\sigma} 
+ c^\dagger_{k_1\sigma} c_{k_2\sigma'} c^\dagger_{p_2\sigma'} c_{p_1\sigma} \big)
\label{V_ex}
\end{eqnarray}
is the exchange interaction term,
\begin{eqnarray}
&& V_{\rm pair} =  \frac{1}{2}\sum_{\sigma\sigma'} \sum_{p_1p_2=1}^{2} \sum_{k_1k_2=1}^{N_{\rm env}} h_{p_1 k_1 p_2 k_2} 
\nonumber\\ &&\times\,
\big(c^\dagger_{p_1\sigma} c^\dagger_{p_2\sigma'} c_{k_2\sigma'} c_{k_1\sigma} 
+\, c^\dagger_{k_1\sigma} c^\dagger_{k_2\sigma'} c_{p_2\sigma'} c_{p_1\sigma} \big)
\label{V_pair}
\end{eqnarray}
describes the hopping of pairs of electrons between the HOMO/LUMO orbitals and the environmnet,
and, finally,  $V_1$ and $V_3$ contain the remaining terms with either one or three summation indexes pointing to the HOMO/LUMO orbitals,
\begin{eqnarray}
&& V_1 = \sum_{\sigma\sigma'} \sum_{p=1}^{2} \sum_{k_1k_2k_3=1}^{N_{\rm env}}  h_{pk_3 k_1k_2} 
\nonumber\\ && \times\,
\left(  c^\dagger_{p\sigma} c^\dagger_{k_1\sigma'} c_{k_2\sigma'} c_{k_3\sigma} +  c^\dagger_{k_3\sigma} c^\dagger_{k_2\sigma'} c_{k_1\sigma'} c_{p\sigma} \right),
\label{V1}
\\
&& V_3 = \sum_{\sigma\sigma'} \sum_{p_1p_2p_3=1}^{2} \sum_{k=1}^{N_{\rm env}}  h_{kp_3p_1p_2} 
\nonumber\\ && \times\,
\left( c^\dagger_{k\sigma} c^\dagger_{p_1\sigma'} c_{p_2\sigma'} c_{p_3\sigma} + c^\dagger_{p_3\sigma} c^\dagger_{p_2\sigma'} c_{p_1\sigma'} c_{k\sigma}  \right).
\label{V3}
\end{eqnarray}

We distinguish two groups of the environment orbitals: the doubly occupied orbitals below the Fermi level, 
which will be denoted by Greek indexes $\alpha,\beta,$
and the empty orbitals above the Fermi level, which we will indicate by the Latin indexes $m,n$. 
We also introduce the total number of the doubly occupied environment orbitals $N_{\rm db}$, which implies that
the number of the empty orbitals is $N_{\rm empt}=N_{\rm env} - N_{\rm db}$.  
Now we perform the averaging of the interaction terms over the environment operators
using the rules $\langle c^\dagger_{m\sigma}c_{n\sigma'}\rangle = 0$, $\langle c^\dagger_{m\sigma}c_{\alpha\sigma'}\rangle = 0$,
$\langle c^\dagger_{\alpha\sigma}c_{\beta\sigma'}\rangle = \delta_{\alpha\beta}\delta_{\sigma\sigma'}$, which apply for non-interacting electrons.
Upon such averaging, the interaction terms $V_t, V_{\rm pair}, V_{1},$ and $V_3$ vanish. 
We sum up all the remaining non-zero terms and add them to the Hamiltonian of the HOMO/LUMO orbitals (\ref{H0}). 
In this way we obtain the "system" Hamiltonian of our model, 
\begin{eqnarray}
H_{\rm sys} = H_{12} + \langle V_{\rm C} \rangle + \langle V_{\rm ex} \rangle.
\end{eqnarray}
It can be written in the form
\begin{eqnarray}
&& H_{\rm sys} = \sum_\sigma\sum_{p_1p_2=1}^{2} \tilde t_{p_1p_2} c^\dagger_{p_1\sigma} c_{p_2\sigma} 
\nonumber\\ &&
+ \frac{1}{2} \sum_{\sigma\sigma'} \sum_{p_1p_2p_3p_4=1}^{2} h_{p_1p_2p_3p_4} c^\dagger_{p_1\sigma} c^\dagger_{p_3\sigma'} c_{p_4\sigma'} c_{p_2\sigma}, 
\label{Hsys}
\end{eqnarray}
where the modified single particle integrals include the usual Hartree-Fock corrections
caused by the influence of the doubly occupied environment orbitals,
\begin{eqnarray}
\tilde t_{p_1p_2} = t_{p_1p_2} + \sum_{k=1}^{N_{\rm db}} ( 2h_{p_1p_2kk} - h_{p_1k p_2k} ).
\label{tilde_t}
\end{eqnarray}
These amplitudes differ from the elements of the Fock matrix because the summation skips the HOMO orbital.

Finally, we re-arrange the terms and write the full Hamiltonian as
\begin{eqnarray}
H = H_{\rm sys} + H_{\rm env} + \tilde V,
\label{H_full}
\end{eqnarray}
where the interaction term now has the form
\begin{eqnarray}
\tilde V = V_t + V_{\rm C} + V_{\rm ex} - \langle V_{\rm C} + V_{\rm ex} \rangle  + V_{\rm pair} + V_1 + V_3.
\label{V_renorm}
\end{eqnarray}

\subsection{System-bath model}
\label{model}

As we show in the Appendices, one can approximately express the full Hamiltonian (\ref{H_full}) in 
the form typical for system-bath models. Specifically, the Hamiltonian of the system (\ref{Hsys}) can be exactly mapped to the form
\begin{eqnarray}
H_{\rm sys} &=&  
\varepsilon_1 + \varepsilon_2 + \frac{U_1 + U_2 + J_{12}}{2} 
+\, \frac{U_1 + U_2 - J_{12}}{2} \sigma_1^z
\nonumber\\ &&
-\, \frac{\Delta_{12}}{2}( 1 + \sigma_1^z ) \otimes \sigma_2^z  + K_{12} \sigma_1^z \otimes \sigma_2^x 
\nonumber\\ &&
+\, \frac{\tilde t_1}{2}    ( \sigma_1^x \otimes \sigma_2^z + \sigma_1^y \otimes \sigma_2^y  +   \sigma_1^x  - \sigma_1^x \otimes \sigma_2^x )
\nonumber\\ &&
+\, \frac{\tilde t_2}{2}  ( \sigma_1^x \otimes \sigma_2^z + \sigma_1^y \otimes \sigma_2^y - \sigma_1^x  + \sigma_1^x \otimes \sigma_2^x ).
\label{H_sys_qubits}
\end{eqnarray} 
Here the two sets of the Pauli matrices $\sigma_1^{\alpha}$, $\sigma_2^{\alpha}$ refer to the operators of the two 
qubits, which have been introduced to span the non-trivial $4\times 4$ subspace of the $6\times 6$ Hilbert space of the Hamiltonian (\ref{Hsys}).
To avoid confusion, we emphasize that these qubit operators differ
from the spin operators of the two electrons occupying the HOMO/LUMO orbitals.
The parameters of the system Hamiltonian (\ref{H_sys_qubits}) are defined in the following way:
we introduce the energies of the two orbitals
\begin{eqnarray}
\varepsilon_1 = \tilde t_{11}, \;\; \varepsilon_2 = \tilde t_{22},
\label{E1E2}
\end{eqnarray}
the charging energies of the HOMO/LUMO orbitals
\begin{eqnarray}
U_1 = {h_{1111}}/{2}, \;\;\; U_2 = {h_{2222}}/{2},
\label{U1U2}
\end{eqnarray} 
the level splitting between the HOMO/LUMO orbitals including Coulomb shifts,
\begin{eqnarray}
\Delta_{12} = \varepsilon_2 + U_2 - \varepsilon_1  - U_1,
\label{Delta_12}
\end{eqnarray}  
the Coulomb repulsion integral between the HOMO and the LUMO orbitals
\begin{eqnarray}
J_{12} = h_{1122},
\label{J_12}
\end{eqnarray}
the exchange integral between them
\begin{eqnarray}
K_{12} = h_{1212},
\label{K_12}
\end{eqnarray}
and the hopping amplitudes
\begin{eqnarray}
\tilde t_1 = \tilde t_{12}   + h_{1112},\;\;
\tilde t_2 = \tilde t_{12}   + h_{1222}.
\label{t1t2}
\end{eqnarray}

Next, we approximately reduce the Hamiltonian of the environment (\ref{Henv}) 
to that of the bath of independent oscillators.
To achieve this, we replace all possible excitations of the environment orbitals, 
which are shown in Fig. \ref{Fig_exc}, by oscillators.
Such replacement is the main point of the random phase approximation \cite{RPA}.
Thus, in our model every electron excitation from a doubly occupied orbital $\alpha$
to an empty orbital $m$, $\alpha\to m$, is represented by an oscillator.  
The oscillators introduced in this way interact with each other. To get
rid of the interaction between them, we numerically
diagonalize the system of coupled oscillators and switch to the 
normal modes. Details of this procedure are presented in the Appendices \ref{Sec_env},\ref{normal}.  
In this way, we bring the environment Hamiltonian to the form
\begin{eqnarray}
H_{\rm env} &=& E^{\rm env}_{\rm HF} +  \sum_{m\alpha} \left( E^{m\alpha}_{\rm corr} - \frac{\hbar\omega_{m\alpha}}{2} \right)
\nonumber\\ &&
+\,  \sum_{m\alpha} \hbar\Omega_{m\alpha}\left( b_{m\alpha}^\dagger b_{m\alpha} + \frac{1}{2} \right).
\label{Henv_diag}
\end{eqnarray}
Here the Hartree-Fock energy of the environment is given by the 
sum over all doubly occupied orbitals $\alpha$ and $\beta$,
\begin{eqnarray}
E_{\rm HF}^{\rm env} = 2\sum_{\alpha=1}^{N_{\rm db}} t_{\alpha\alpha}  
+ \sum_{\alpha,\beta=1}^{N_{\rm db}} ( 2h_{\alpha\alpha\beta\beta} - h_{\alpha\beta\alpha\beta} ),
\label{EHF}
\end{eqnarray} 
$b_{m\alpha},b^\dagger_{m\alpha}$ are the ladder operators of the normal
modes of the environment, with each of them approximately corresponding to a transition $\alpha\to m$
with some admixture of other transitions,
$\omega_{m\alpha}$ are the frequencies of the uncoupled transitions in the environment,
$\Omega_{m\alpha}$ are the frequencies of the normal modes, and $E^{m\alpha}_{\rm corr}$ are the
correlation energies associated with the transitions $\alpha\to m$. 
The frequencies $\omega_{m\alpha}$ and the correlation energies $E^{m\alpha}_{\rm corr}$ \cite{Szabo}
for the transitions from the doubly occupied to the empty orbitals,
represented by long arrows in the left side of both panels of Fig. \ref{Fig_exc},
have the form
\begin{eqnarray}
\hbar\omega_{m\alpha} &=& \sqrt{\Delta_{m\alpha}^2 + K_{m\alpha}^2} - U_\alpha - U_m + J_{m\alpha} + K_{m\alpha},
\nonumber\\
\label{omega}\\
E^{m\alpha}_{\rm corr} &=& \Delta_{m\alpha} - \sqrt{\Delta_{m\alpha}^2 + K_{m\alpha}^2}.
\label{E_corr}
\end{eqnarray}
Here we have defined the parameters of the transitions in the same way
as for the HOMO/LUMO system,
\begin{eqnarray}
&& U_m = h_{mmmm}/2, \;\;\; U_{\alpha} = h_{\alpha\alpha\alpha\alpha}/2,
\nonumber\\ &&
J_{m\alpha} = h_{mm\alpha \alpha}, \;\;\; K_{m\alpha} = h_{m\alpha m\alpha}, 
\label{param}
\end{eqnarray}
and $\Delta_{m\alpha}$ is the energy difference
between the orbitals $m$ and $\alpha$ modified by Coulomb interaction. 
To express this parameter in terms of previously defined values, we introduce the 
Hartree-Fock energies of the environment orbitals
\begin{eqnarray}
\varepsilon_k = t_{kk} + \sum_{\alpha=1}^{N_{\rm db}} (2h_{kk\alpha\alpha} - h_{k\alpha k\alpha}).
\label{varepsilon}
\end{eqnarray}
This definition applies to both doubly occupied and empty orbitals. The energy difference
$\Delta_{m\alpha}$ is then defined as
\begin{eqnarray}
\Delta_{m\alpha} = \varepsilon_m - \varepsilon_\alpha + U_m + U_\alpha - 2J_{m\alpha} + K_{m\alpha}.
\label{Dmalpha}
\end{eqnarray}
The frequencies $\hbar\omega_{m\alpha}$ and the level splittings $\Delta_{m\alpha}$ for
the transitions involving HOMO/LUMO orbitals, shown by vertical arrows in the right side of both panels
of Fig. \ref{Fig_exc}, are given in the Appendix, see Eqs. (\ref{omega_m1},\ref{Delta_m1},\ref{omega_2a},\ref{Delta_2a}) 
for the molecules with the singlet ground state of the system Hamiltonian and Eqs. (\ref{omega_ij}) for the molecules with
the triplet ground state of $H_{\rm sys}$.
To obtain the frequencies of the normal modes $\Omega_{m\alpha}$, one needs to diagonalize the matrix $\hat M$, which
is derived from the Hamiltonian of the coupled bath oscillators and fully determines the properties
of the bath in the RPA approximation. 
For a molecule with the singlet ground state of the system Hamiltonian $H_{\rm sys}$ this matrix
has the following matrix elements
\begin{eqnarray}
&& \hat M_{m\alpha,n\beta} = \omega^2_{m\alpha}\delta_{m\alpha,n\beta} 
\nonumber\\ &&
+\, \frac{4\sqrt{\omega_{m\alpha}\omega_{n\beta}}}{\hbar} r_{m\alpha} h_{m\alpha,n\beta} r_{n\beta} [ 1 - \delta_{m\alpha,n\beta} ]. 
\label{M}
\end{eqnarray}
Here the parameters $r_{m\alpha}$ are defined as
\begin{eqnarray}
r_{m\alpha} = \frac{ \sqrt{\Delta_{m\alpha}^2 + K_{m\alpha}^2} + \Delta_{m\alpha} - K_{m\alpha}}
{\sqrt{ K_{m\alpha}^2 + \left( \sqrt{\Delta_{m\alpha}^2 + K_{m\alpha}^2} + \Delta_{m\alpha} \right)^2 }}.
\label{r}
\end{eqnarray}
For molecules with the triplet ground state of $H_{\rm sys}$ the last term of the matrix (\ref{M}), describing
the potential energy of the interacting environment oscillators, should be modified in accordance with the form of
the Hamiltonian (\ref{H_env_triplet}) given in Appendix \ref{Sec_H_trilpet}.
The normal mode frequencies $\Omega_{m\alpha}$ are obtained from the equation
\begin{eqnarray}
\det[\hat M_{m\alpha,n\beta} - \Omega^2 \delta_{m\alpha,n\beta}] = 0.
\label{Omega_ma}
\end{eqnarray} 
Each pair of indexes $m\alpha$ or $n\beta$ in Eqs. (\ref{M},\ref{Omega_ma}) should be viewed as a single index.
Thus, the matrix $\hat M$ has the dimensions $N_{\rm RPA}\times N_{\rm RPA}$, where the total number of
the RPA oscillators in the environment, $N_{\rm RPA}$, equals to the number of possible pairs $m\alpha$. 
For a molecule with the singlet ground state of $H_{\rm sys}$ this number is given by
\begin{eqnarray}
N_{\rm RPA} = (N_{\rm db} +1) \times (N_{\rm empt} + 1) - 1,
\label{N_RPA_sg}
\end{eqnarray}
and for a molecule with the triplet ground state of the system Hamiltonian the total size of the environment is
\begin{eqnarray}
N_{\rm RPA} = (N_{\rm db} +2) \times (N_{\rm empt} + 2) - 4.
\label{N_RPA_tr}
\end{eqnarray}
Here, as before, $N_{\rm db}$ is the number of the doubly occupied orbitals, $N_{\rm empt}$ is the number of the empty orbitals
so that the total number of orbitals is $N = N_{\rm db} + N_{\rm empt} + 2$ with the number 2 accounting for the HOMO/LUMO orbitals.
The physical meaning of the frequencies $\Omega_{m\alpha}$ is the following: $\hbar\Omega_{m\alpha}$
are approximately equal to the vertical excitation energies of the singlet states of the molecule if it
has the singlet ground state, and $\hbar\Omega_{m\alpha}$ are
the vertical excitation energies of the triplet states if the molecule has the triplet ground state. 
  
Finally, the part of the Hamiltonian responsible for the interaction between the HOMO/LUMO orbitals and the environment (\ref{Vint}) 
can be approximated as
\begin{eqnarray}
\tilde V &=&  \sum_{m\alpha}  ( b^\dagger_{m\alpha} + b_{m\alpha} ) 
\bigg[ g_{12}^{m\alpha} ( \sigma_1^x \otimes \sigma_2^z + \sigma_1^y \otimes \sigma_2^y  )
\nonumber\\ &&   
+\, g_{11}^{m\alpha} + g_{22}^{m\alpha} + \frac{g_{11}^{m\alpha} - g_{22}^{m\alpha}}{2} ( 1 + \sigma_1^z ) \otimes \sigma_2^z  \bigg]
\nonumber\\ &&
+\, \sum_{q=1}^2 \sum_{m\alpha}  
\bigg[ \frac{g_{1q}^{m\alpha} g_{q1}^{m\alpha} + g_{2q}^{n\beta} g_{q2}^{n\beta}}{\hbar\Omega_{n\beta}}
\nonumber\\ &&   
+\, \frac{g_{1 q}^{m\alpha} g_{q2}^{m\alpha}}{\hbar\Omega_{n\beta}} ( \sigma_1^x \otimes \sigma_2^z + \sigma_1^y \otimes \sigma_2^y  )
\nonumber\\ &&
+\, \frac{g_{1q}^{m\alpha} g_{q1}^{m\alpha} - g_{2q}^{m\alpha} g_{q2}^{m\alpha}}{2\hbar\Omega_{n\beta}} ( 1 + \sigma_1^z ) \otimes \sigma_2^z  \bigg].
\label{V_int_normal}
\end{eqnarray}
The coupling constants between the system and the environment are 
\begin{eqnarray}
g_{p_1p_2}^{n\beta} = \sum_{m\alpha} \sqrt{\frac{2\omega_{m\alpha}}{\Omega_{n\beta}}} h_{p_1p_2 m\alpha} r_{m\alpha}  \hat S_{m\alpha,n\beta},
\label{g}
\end{eqnarray}
where the matrix with the elements $\hat S_{m\alpha,n\beta}$ is composed of the column eigenvectors of the matrix $\hat M$ (\ref{M}). 
The physical meaning of various interaction terms in Eq. (\ref{V_int_normal}) is the following: 
the terms containing $g_{12}^{m\alpha}$ describe the fluctuations of the hopping amplitudes between the HOMO/LUMO orbitals,
which may be approximately expressed as $\tilde t_{1,2} + g_{12}^{m\alpha}( b^\dagger_{m\alpha} + b_{m\alpha} )$;  
likewise the terms containing $g_{11}^{m\alpha} - g_{22}^{m\alpha}$ describe the fluctuations of the energy splitting
between the HOMO/LUMO orbitals, which may be expressed as $\Delta_ {12} - (g_{11}^{m\alpha} - g_{22}^{m\alpha})( b^\dagger_{m\alpha} + b_{m\alpha} ) $;
and, finally, the last three lines of the interaction Hamiltonian (\ref{V_int_normal}) constitute the counter-term, which restores
the normal ordering of the fermionic operators in the Lamb shift, see Appendix \ref{counter}
for details.  

From the Hamiltonian (\ref{Henv_diag}) we obtain the ground state energy of the environment in the form
\begin{eqnarray}
E_{\rm gr}^{\rm env} = E_{\rm HF}^{\rm env} + \sum_{m\alpha} \left( E_{m\alpha}^{\rm corr} + \frac{\hbar(\Omega_{m\alpha} - \omega_{m\alpha})}{2}\right).
\end{eqnarray}
It coincides with the energy obtained in the RPA approximation \cite{Furche,RPA}, as expected.

The Hamiltonian (\ref{H_sys_qubits},\ref{Henv_diag},\ref{V_int_normal}) is similar to that considered 
by Scott and Booth in Ref. \cite{Scott1} and earlier by us in Ref. \cite{RPA}. 
Here we have introduced the following novel features compared to Ref. \cite{RPA}: (i) we have added new oscillators
to the environment, which describe the transitions between the HOMO/LUMO and the environment orbitals, 
(ii) the frequencies of the environment oscillators (\ref{omega}) are evaluated more accurately than in Ref. \cite{RPA},
where simple Hartee-Fock approximation has been used for this purpose, and (iii) we have introduced the
counter-term in the interaction Hamiltonian, which allows us to go beyond the static screening model.

\section{Hamiltonian of a molecule in terms of qubit operators}
\label{qubit_env}

In this section we present our main result expressing the Hamiltonian of a molecule
in terms of qubit operators. We will use this form of the Hamiltonian for subsequent numerical simulations.

For the majority of molecules the interaction constants $g_{p_1p_2}^{m\alpha}$ are very small and the oscillators
of the environment are only weakly excited. Therefore the oscillators can be replaced by two level systems, or qubits, without significant loss
of accuracy. Hence, one can express the Hamiltonian of a molecule (\ref{H}) in the approximate form
\begin{eqnarray}
H = H_{\rm sys} + H_{\rm env} + \tilde V,
\label{H_qubit_env}
\end{eqnarray}
where the system Hamiltonian is given by Eq. (\ref{H_sys_qubits}) as before, 
\begin{eqnarray}
H_{\rm sys} &=&  
\varepsilon_1 + \varepsilon_2 + \frac{U_1 + U_2 + J_{12}}{2} 
+\, \frac{U_1 + U_2 - J_{12}}{2} \sigma_1^z
\nonumber\\ &&
-\, \frac{\Delta_{12}}{2}( 1 + \sigma_1^z ) \otimes \sigma_2^z  + K_{12} \sigma_1^z \otimes \sigma_2^x 
\nonumber\\ &&
+\, \frac{\tilde t_1}{2}    ( \sigma_1^x \otimes \sigma_2^z + \sigma_1^y \otimes \sigma_2^y  +   \sigma_1^x  - \sigma_1^x \otimes \sigma_2^x )
\nonumber\\ &&
+\, \frac{\tilde t_2}{2}  ( \sigma_1^x \otimes \sigma_2^z + \sigma_1^y \otimes \sigma_2^y - \sigma_1^x  + \sigma_1^x \otimes \sigma_2^x ),
\label{H_sys_qubits_2}
\end{eqnarray} 
the environment Hamiltonian
has the form
\begin{eqnarray}
H_{\rm env} &=& E^{\rm env}_{\rm HF} +  \sum_{m\alpha} \left[ E^{m\alpha}_{\rm corr} - \frac{\hbar\omega_{m\alpha}}{2} \right]
\nonumber\\ &&
+\,  \hbar\Omega_{m\alpha}\left( \sigma^+_{m\alpha} \sigma^-_{m\alpha} + \frac{1}{2} \right),
\label{H_env_qubits}
\end{eqnarray}
and the interaction term becomes
\begin{eqnarray}
\tilde V &=&  \sum_{m\alpha}  \sigma^x_{m\alpha}  
\bigg[ g_{12}^{m\alpha} ( \sigma_1^x \otimes \sigma_2^z + \sigma_1^y \otimes \sigma_2^y  )
\nonumber\\ &&   
+\, g_{11}^{m\alpha} + g_{22}^{m\alpha} + \frac{g_{11}^{m\alpha} - g_{22}^{m\alpha}}{2} ( 1 + \sigma_1^z ) \otimes \sigma_2^z  \bigg]
\nonumber\\ &&
+\, \sum_{q=1}^2 \sum_{m\alpha}  
\bigg[ \frac{g_{1q}^{m\alpha} g_{q1}^{m\alpha} + g_{2q}^{m\alpha} g_{q2}^{m\alpha}}{\hbar\Omega_{m\alpha}}
\nonumber\\ &&   
+\, \frac{g_{1 q}^{m\alpha} g_{q2}^{m\alpha}}{\hbar\Omega_{m\alpha}} ( \sigma_1^x \otimes \sigma_2^z + \sigma_1^y \otimes \sigma_2^y  )
\nonumber\\ &&
+\, \frac{g_{1q}^{m\alpha} g_{q1}^{m\alpha} - g_{2q}^{m\alpha} g_{q2}^{m\alpha}}{2\hbar\Omega_{m\alpha}} ( 1 + \sigma_1^z ) \otimes \sigma_2^z  \bigg].
\label{V_int_qubits}
\end{eqnarray}
Here the Pauli matrices $\sigma^x_{m\alpha},$ $\sigma^\pm_{m\alpha} = ( \sigma_{m\alpha} \pm i \sigma_{m\alpha}^y )/2$ describe the
environment qubits, which replace the environment oscillators considered in the previous section.

\section{Static screening approximation}
\label{Sec_static}

Accurate quantum chemistry simulations require keeping very large number of oscillators
or qubits in the environment. To achieve chemical accuracy even for a relatively small molecule, 
one typically needs to retain from 2000 to 10000 qubits. Numerically solving the Hamiltonian
(\ref{H_qubit_env}) with so many qubits is quite challenging. 
Fortunately, most of the coupling constants (\ref{g}) are very small, 
which allows one to use the perturbation theory. Such perturbation theory leads to the static screening approximation
proposed in Ref. \cite{RPA}. The basic idea behind this approximation is simple: the electron repulsion integrals,
which determine the system parameters (\ref{U1U2},\ref{Delta_12},\ref{J_12},\ref{K_12}), 
should be modified by replacing the unscreened Coulomb potential in the corresponding integrals (\ref{eri}) with its screened
version. For example, in the simplest case of homogeneous electron gas this amounts to the following replacement: $e^2/r \to e^2 \exp(-r/\lambda_{\rm TF})/r$,
where $\lambda_{\rm TF}$ is the Thomas-Fermi screening length. The static screening approximation should work best
for diradical molecules, for which the energy splitting between the HOMO and the LUMO orbitals is small compared
to the frequencies of the environment oscillators, $\Delta_{12} \ll \hbar\Omega_{m\alpha}$.
However, this approximation works reasonably well even for molecules with $\Delta_{12} \sim \hbar\Omega_{m\alpha}$.

Within the static screening approximation the Hamiltonian of a molecule takes the form
\begin{eqnarray}
H = \tilde H_{\rm sys} + H_{\rm env},
\label{H_static}
\end{eqnarray}
i.e. the system and the environment become decoupled. The Hamiltonian of the environment retains 
its original form (\ref{H_env_qubits}), while the parameters of the system Hamiltonian get renormalized.
After such renormalization, the system Hamiltonian takes the form
\begin{eqnarray}
\tilde H_{\rm sys} &=&  
\varepsilon_1 + \varepsilon_2 + \frac{\tilde U_1 + \tilde U_2 + \tilde J_{12}}{2} 
+\, \frac{\tilde U_1 + \tilde U_2 - \tilde J_{12}}{2} \sigma_1^z
\nonumber\\ &&
-\, \frac{\varepsilon_2 + \tilde U_2 - \varepsilon_1  - \tilde U_1}{2}( 1 + \sigma_1^z ) \otimes \sigma_2^z  + \tilde K_{12} \sigma_1^z \otimes \sigma_2^x 
\nonumber\\ &&
+\, \frac{\tilde \tau_1}{2}    ( \sigma_1^x \otimes \sigma_2^z + \sigma_1^y \otimes \sigma_2^y  +   \sigma_1^x  - \sigma_1^x \otimes \sigma_2^x )
\nonumber\\ &&
+\, \frac{\tilde \tau_2}{2}  ( \sigma_1^x \otimes \sigma_2^z + \sigma_1^y \otimes \sigma_2^y - \sigma_1^x  + \sigma_1^x \otimes \sigma_2^x ),
\label{H_sys_static}
\end{eqnarray}    
where the renormalized parameters are
\begin{eqnarray}
\tilde U_j &=& U_j - \sum_{k=1}^{N_{\rm RPA}}\frac{(g^{(k)}_{jj})^2}{\hbar\Omega_k},   
\nonumber\\ 
\tilde J_{12} &=& J_{12} - \sum_{k=1}^{N_{\rm RPA}}\frac{ 2 g^{(k)}_{11} g^{(k)}_{22} }{\hbar\Omega_k}, 
\nonumber\\ 
\tilde K_{12} &=& K_{12} - \sum_{k=1}^{N_{\rm RPA}}\frac{2(g^{(k)}_{12})^2}{\hbar\Omega_k},
\nonumber\\ 
\tilde \tau_j &=& \tilde t_j - \sum_{k=1}^{N_{\rm RPA}}\frac{ 2g^{(k)}_{12} g^{(k)}_{jj} }{\hbar\Omega_k}.
\label{param_static}
\end{eqnarray}
Here in the sums we have replaced each pair of indexes $m\alpha$, referring to the transition $\alpha\to m$, by a single index $k$,
which enumerates all normal modes of the environment and the corresponding qubits.
The details of the derivation are presented in Appendix F, the parameters (\ref{param_static}) are obtained
by comparing the original form of the system Hamiltonian (\ref{Hsys}) with the renormalized form (\ref{H_static_2}).
Diagonalizing the renormalized Hamiltonian (\ref{H_sys_static}), which is a $4\times 4$ matrix,
one can easily obtain the lowest excitation energy of a molecule. 
To obtain higher excitation energies, one should diagonalize the full Hamiltonian (\ref{H_static}),
which can be easily done because the system and the environment are decoupled. 
In particular, the lowest vertical excitation energies of the singlet states are equal to the frequencies of the normal
modes, $\Delta E_{\rm sg}^{(k)} = \hbar\Omega_{k}$, if the ground state of a molecule
is a singlet. For triplet molecules the frequencies $\hbar\Omega_k$
correspond to the lowest triplet excitations.

\section{Mixed approximation}
\label{mixed}

Although the static approximation described in the previous section already provides rather
good accuracy for many molecules, one should go beyond it if one wants to achieve 
the chemical accuracy of 1 kcal/mol = 0.043 eV.
To approach such accuracy without
running into computational problems, we introduce mixed approximation. 
Namely, we treat a manageable number of the strongest coupled qubits of the environment exactly 
and apply the static screening approximation to all remaining environment qubits. 
To measure the coupling strength, for every
environment qubit we calculate the parameter
\begin{eqnarray}
L_k = \frac{\left(g^{(k)}_{12}\right)^2 + \left(g^{(k)}_{11} - g^{(k)}_{22}\right)^2}{\hbar\Omega_k}.
\end{eqnarray} 
We rank the qubits according to the value of $L_k$
and select $N_q$ strongest coupled qubits with the largest $L_k$. 
Next, we approximate the Hamiltonian (\ref{H_qubit_env}) as
\begin{eqnarray}
H = \tilde H_{\rm sys} + H_{\rm env} + \tilde V,
\label{H_Nq}
\end{eqnarray}
where the system Hamiltonian $\tilde H_{\rm sys}$ has the form (\ref{H_sys_static})
with the renormalized parameters determined only by weakly coupled oscillators with the indexes in the range $N_q+1 \leq k \leq N_{\rm RPA}$, 
\begin{eqnarray}
\tilde U_j &=& U_j - \sum_{k=N_q+1}^{N_{\rm RPA}}\frac{(g^{(k)}_{jj})^2}{\hbar\Omega_k},  
\nonumber\\ 
\tilde J_{12} &=& J_{12} - \sum_{k=N_q+1}^{N_{\rm RPA}}\frac{ 2 g^{(k)}_{11} g^{(k)}_{22} }{\hbar\Omega_k}, \;\;
\nonumber\\
\tilde K_{12} &=& K_{12} - \sum_{k=N_q+1}^{N_{\rm RPA}}\frac{2(g^{(k)}_{12})^2}{\hbar\Omega_k},
\nonumber\\ 
\tilde \tau_j &=& \tilde t_j - \sum_{k=N_q+1}^{N_{\rm RPA}}\frac{ 2g^{(k)}_{12} g^{(k)}_{jj} }{\hbar\Omega_k}.
\label{params_2}
\end{eqnarray}
The environment Hamiltonian has the unchanged form (\ref{H_env_qubits})
and the interaction term includes only $N_q$ strongest coupled environment qubits,
\begin{eqnarray}
\tilde V &=&  \sum_{k=1}^{N_q}  \sigma^x_{k} \bigg[ g_{12}^{(k)} ( \sigma_1^x \otimes \sigma_2^z + \sigma_1^y \otimes \sigma_2^y  )
\nonumber\\ &&   
+\, g_{11}^{(k)} + g_{22}^{(k)} + \frac{g_{11}^{(k)} - g_{22}^{(k)}}{2} ( 1 + \sigma_1^z ) \otimes \sigma_2^z  \bigg]
\nonumber\\ &&
+\, \sum_{k=1}^{N_q}\sum_{q=1}^2   
\bigg[ \frac{g_{1q}^{(k)} g_{q1}^{(k)} + g_{2q}^{(k)} g_{q2}^{(k)}}{\hbar\Omega_{k}}
\nonumber\\ &&   
+\, \frac{g_{1 q}^{(k)} g_{q2}^{(k)}}{\hbar\Omega_{k}} ( \sigma_1^x \otimes \sigma_2^z + \sigma_1^y \otimes \sigma_2^y  )
\nonumber\\ &&
+\, \frac{g_{1q}^{(k)} g_{q1}^{(k)} - g_{2q}^{(k)} g_{q2}^{(k)}}{2\hbar\Omega_{k}} ( 1 + \sigma_1^z ) \otimes \sigma_2^z  \bigg].
\label{V_int_Nq}
\end{eqnarray}

\section{Results and discussion}
\label{results}

\begin{table*}
\footnotesize
\begin{tabular}{|c|c|c|c|c|c|c|c|c|c|c|c|c|c|c|}
\hline
                                & 1       &  2       &  3       &  4      &  5        &  6       & 7         & 8        &  9         & 10       &  11   &  12        &  13       &       \\
\hline
Molecule                        & CPC     &  MBQ     &  OXA     &  TMM    &  CH$_2$   &$m$-Xyl   & TOTMB     & $p$-benz &  $o$-benz  & $m$-benz &  TME  &  PN        &  PC       &  AAE$^*$  \\
                                &         &          &          &         &           &          &           &          &            &          &       &            &           &       \\
\hline
model of env.                   & tr      &  tr      &  sg      &  tr     &  tr       & tr       &  sg       &  tr      &  tr        &  sg      & sg    &  tr        &  tr       &       \\
\hline
$\Delta E_{\rm ST}$,            & 14.71   & 6.03     & -7.89    & 33.50   & 16.59     & 2.61     & -6.34     & -4.41    & -57.64     & -29.46   & -3.29 & 22.37      & 5.13      & 5.28   \\
static app.                     &         &          &          &         &           &          &           &          &            &          &       &            &           &        \\
\hline
$\Delta E_{\rm ST}$,            & 15.14   & 6.26     & -6.51    & 35.87   & 17.92     & 3.00     & -5.83     & -3.21    & -54.14     & -25.11   & -3.04 & 22.59      & 5.25      & 4.68   \\
env. of 62 qubits               &         &          &          &         &           &          &           &          &            &          &       &            &           &        \\
\hline
$\Delta E_{\rm ST}$, MkCCSD(T)  & 13.12   &  9.86    & -2.21    &  23.03  &  19.57    &  13.16   &  -3.69    &  -4.82   &  -52.63    &  -24.28  & -5.18 & 31.84      &  14.02    & \dots    \\
{\bf Reference}                 &         &          &          &         &           &          &           &          &            &          &       &            &           &        \\
\hline
\end{tabular}
\caption{Singlet-triplet gaps $\Delta E_{\rm ST} = E_{\rm S}-E_{\rm T}$ (kcal/mol) for 13 diradical molecules. 
Model of the environment "sg" is applied to the molecules with the singlet ground state of the system Hamiltonian $H_{\rm sys}$ (\ref{H_sys_qubits}), 
while the "triplet" environment model, "tr", is applied to the molecules with the triplet ground states of $H_{\rm sys}$. \\
$^*$ AAE stands for the absolute average error.}
\label{table23}
\end{table*}

In this section we test the proposed model on a number of molecules.
As a starting point, for each molecule we perform the restricted open-shell Hartree-Fock (ROHF) simulation
with the {\bf PySCF} python package \cite{Sun1,Sun2}. We have simulated 
the triplet states of the molecules, in which both HOMO and LUMO orbitals are populated by one
electron. Such states are known to be well approximated by a single Slater determinant.
This ensures better quality of the input parameters of our model, namely, 
of the single electron matrix elements $t_{pq}$ (\ref{t_pq}) 
and of the Coulomb repulsion integrals $h_{pqrs}$ (\ref{eri}), which we
store for subsequent analysis.

To begin with, we have tested our model on the set of 13 diradical molecules, which we
have earlier considered in Ref. \cite{RPA}. 
These molecules are: 
(1) cyclopentane cation  (CPC), (2) $m$-benzoquinone (MBQ), (3) oxyallyl (OXA), (4) trimethylenemethane (TMM),
(5) methylene (CH$_2$), (6) $m$-xylylene ($m$-XYL), (7) tetraoxatetramethylenebenzene (TOTMB), (8) $p-$benzyne,
(9) $o-$benzyne, (10) $m-$benzyne, (11) tetramethyleneethylene (TME), (12) phenylnitrene (PN) and (13) phenylcarbene (PC). 
For all these molecules we have computed the vertical singlet-triplet gaps, defined as the difference between the energies of
the singlet and of the triplet states, $\Delta E_{\rm ST} = E_{\rm S}-E_{\rm T}$, in three different ways.
First, we have applied the static screening approximation and evaluated 
$\Delta E_{\rm ST}$ for every molecule by 
diagonalizing the renormalized system Hamiltonian (\ref{H_sys_static}).
Subsequently, we have tested the mixed approximation and diagonalized 
the Hamiltonian (\ref{H_Nq}) retaining 62 qubits in the bath.
To perform this diagonalization, we have used the Quantum Solver python package developed in our
company. It allows one to project the Hamiltonian (\ref{H_Nq}) from its full $2^{64} \times 2^{64}$ Hilbert space
to a much more compact subspace with the number of spin excitations
not exceeding certain reasonably small value, and then
to diagonalize the projected Hamiltonian. 
We have verified that allowing no more than 4 spin excitations is sufficient to achieve the
convergence for the few lowest eigenenergies of the Hamiltonian.
In the quantum chemistry language this projection corresponds to allowing no more than quadruple excitations 
as it is done, for example, in the coupled-electron pair (CEPA) methods \cite{Neese}.
Finally, we have compared the results to the reference values obtained with the high accuracy quantum
chemistry simulations based on    
the Multi-reference state-specific Mukherjee’s
coupled cluster method with non-iterative triple excitations (MkCCSD(T)) \cite{MKCCSD1,MKCCSD2}.
We have used the same def2-TZVP basis set for all these simulations.
The results are summarized in Table \ref{table23}.
For some molecules the system Hamiltonian $H_{\rm sys}$ (\ref{H_sys_qubits_2})
has the singlet ground state, they are indicated by the symbol "sg" in the first line of the Table  \ref{table23}.
For the remaining molecules, indicated by "tr", $H_{\rm sys}$ has the triplet ground state.
The average absolute error (AAE) of the static approximation turned out to be 
$5.28$ kcal/mol (0.229 eV). By keeping 
62 qubits  in the bath  we have reduced the error to 4.68 kcal/mol (0.203 eV).
Thus, going beyond the static model results in noticeable improvement of the results.
The chemical accuracy  of 1 kcal/mol (0.043 eV) has been achieved for one molecule --- $m$-benzene,
and for three other molecules -- methylene (CH$_2$), $p$-benzene and $o$-benzene  -- the error has been below 2 kcal/mol.

Next, we have tested the performance of the static approximation on a set of 20 closed shell molecules with
relatively large excitation energies, i.e. on the molecules which are not diradicals. 
The molecules have been chosen from the QUEST database \cite{QUEST},
which has been developed for benchmarking of quantum chemistry methods.
The database contains optimized geometries and vertical excitation energies, evaluated with
advanced quantum chemistry methods, for more than 80 molecules. 
We have selected 20 molecules from this list, for which the restricted Hartree-Fock simulation well converged
and for which our model provided reasonable outcomes. 
The results of the simulations are presented in Table \ref{table2} of the Appendix.
For each molecule we have computed vertical
excitation energies of the lowest triplet and of the lowest singlet
states applying the static approximation, i.e. by diagonalizing the renormalized uncoupled Hamiltonian (\ref{H_static}).
We have used triple-$\zeta$ quality basis set def2-TZVPD \cite{basis_1,basis_2,basis_3}
for these simulations. As in the case of diradicals,  
the molecules can be split in the two groups: those with the singlet ground state of the system Hamiltonian $H_{\rm sys}$ (\ref{H_sys_qubits_2}),
they are indicated by the symbol "sg" in the last column of the Table  \ref{table2},
and those with the triplet ground state of $H_{\rm sys}$ indicated by "tr" symbol.
We have used the corresponding models of the environment for these molecules.
However, after the interaction with the environment is accounted for, 
the ground state for all 20 molecules becomes a singlet, which is 
the well known property of these molecules.  
For all molecules, the excitation energy of the lowest triplet state is given by 
the difference between the energies of the first excited and of the ground state of the
renormalized system Hamiltonian $\tilde H_{\rm sys}$ (\ref{H_sys_static}).
As for the lowest singlet excitation energy,
it equals to the lowest frequency of the bath qubits, $\min_{k}[ \hbar\Omega_{k} ]$,
if the  Hamiltonian $H_{\rm sys}$ (\ref{H_sys_qubits_2}) has the singlet ground state, 
and it equals to the gap between the second excited and the ground state of the renormalized 
Hamiltonian (\ref{H_sys_static}) if $H_{\rm sys}$ has the triplet ground state.
Comparing our results to the reference values from the QUEST database, we
obtain the average error of 0.28 eV for the lowest triplet states and 0.27 eV for the lowest singlet states.
These errors are just a little worse than those for the diradicals listed in Table \ref{table23}.
For comparison, we have also 
performed the standard density functional theory (DFT) simulations for the triplet excitations and 
the time-dependent density functional theory (TD-DFT) simulations for the singlet excitations with the B3LYP functional. 
The resulting average absolute errors have been found to be 0.17 eV for
the lowest triplets and 0.26 eV for the lowest singlets, i.e. the DFT results are
a bit better than those of the static screening approximation.

\begin{table}[!ht]
\begin{tabular}{|l|l||c|c|c|c|}
\hline
     & Molecule              & QUEST      & stat. app. & 62 qubits  & DFT        \\
\hline
\hline
 1   & Benzene               & 4.16       & 4.451      & 4.311      & 4.595      \\
\hline
 2   & Butadiene             & 3.36       & 3.018      & 2.937      & 3.304      \\
\hline
 3   & Cyclopentadiene       & 3.31       & 3.359      & 3.283      & 3.243      \\
\hline
 4   & Diacetylene           & 4.10       & 4.326      & 4.048      & 4.595      \\
\hline
 5   & Furan                 & 4.20       & 4.117      & 4.020      & 4.100      \\
\hline
 6   & Hexatriene            & 2.73       & 2.583      & 2.537      & 2.678      \\
\hline
 7   & Imidazole             & 4.73       & 4.889      & 4.790      & 4.696      \\
\hline
 8   & Pyrrole               & 4.51       & 4.575      & 4.470      & 4.486      \\
\hline
 9   & Thiophene             & 3.97       & 4.018      & 3.919      & 3.905      \\
\hline
     & AAE                   &            & 0.157      & 0.131      & 0.148      \\
\hline
\end{tabular}
\caption{Vertical excitation energies of the lowest TRIPLET states (eV)
for 9 molecules with $g_{11}^{(k)} \approx g_{22}^{(k)}\approx 0$.}
\label{Table_tr}
\end{table}

\begin{figure*}
\begin{tabular}{ccc}
\includegraphics[width=6cm]{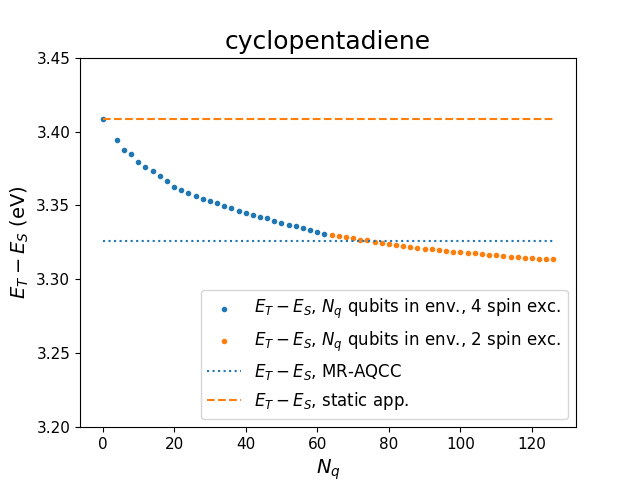}  & 
\includegraphics[width=6cm]{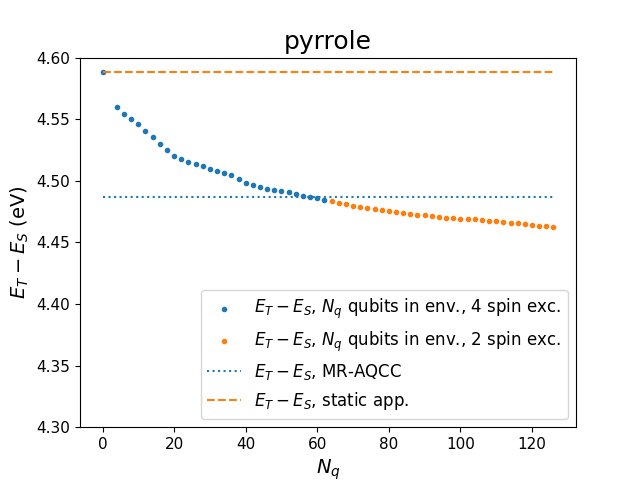}          &
\includegraphics[width=6cm]{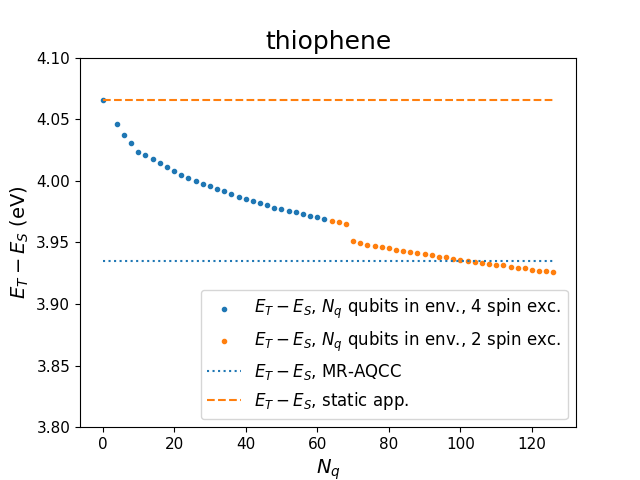}
\end{tabular}
\caption{Vertical triplet-singlet gaps $E_{\rm T}-E_{\rm S}$ (eV) for three selected molecules:
cyclopentadiene with the total number of environment qubits $N_{\rm RPA}=3977$; 
pyrrole with the total number of environment qubits $N_{\rm RPA}=3815$; 
and thiophene with the total number of environment qubits $N_{\rm RPA}=4575$. 
Blue dots indicate the results of the Quantum Solver simulations with the number of qubits in the environment in the range $0\leq N_q \leq 62$
and with no more than 4 spin excitations, orange dots show the $E_{\rm T}-E_{\rm S}$ gaps computed for $63\leq N_q \leq 126$ and
with no more than 2 spin excitations, orange dashed  
lines show the values obtained within the static approximation of Sec. \ref{Sec_static}, and blue dotted lines are the MR-AQCC values.
Blue and orange dots are computed within the mixed model of Sec. \ref{mixed}, in which $N_q$ strongest coupled bath qubits are
kept in the environment and the remaining $N_{\rm RPA}-N_q$ weaker coupled bath qubits are absorbed in the Lamb shift,
which results in the renormalization of the system parameters (\ref{params_2}).}
\label{dETS}
\end{figure*}

Closer look at the results listed in Tables \ref{table23},\ref{table2} reveals that the
molecules with the singlet ground state of $H_{\rm sys}$ are consistently
better described by our model. To look into this deeper,
we have selected 9 molecules with the singlet ground state of 
$H_{\rm sys}$ (\ref{H_sys_qubits_2}) and, in addition, 
with very small longitudinal coupling constants, $g_{11}^{(k)}, g_{22}^{(k)}\approx 0$,
for all or most of the 62 strongest coupled bath qubits.
We have diagonalized the mixed Hamiltonian (\ref{H_Nq}-\ref{V_int_Nq}) for these molecules 
with the Quantum Solver keeping 62 qubits in the environment and allowing no more than 4 spin excitations. 
We have also applied static screening approximation, DFT and TD-DFT 
and computed the lowest triplets and singlets with all these methods. 
The results are presented in Tables \ref{Table_tr},\ref{Table_sg}.
All three methods result in similar errors of $\sim 0.15$ eV for the
triplet excitations and $\sim 0.2 - 0.3$ eV for the singlet ones. 
The simulation with the 62 bath qubits is
slightly more accurate for the lowest triplet excitations.
Chemical accuracy for the lowest triplets has been achieved for three molecules out of nine.
These are cyclopentadiene, pyrrole and thiophene. The errors for the lowest singlets
are slightly higher.

\begin{table}[!ht]
\begin{tabular}{|l|l||c|c|c|c|}
\hline
     & Molecule              & QUEST      & stat. app.  & 62 qubits  & TD-DFT     \\
\hline
\hline
 1   & Benzene               & 5.06       & 5.435       & 5.435      & 5.437      \\
\hline
 2   & Butadiene             & 6.22       & 5.735       & 5.735      & 5.686      \\
\hline
 3   & Cyclopentadiene       & 5.54       & 5.498       & 5.498      & 5.008      \\
\hline
 4   & Diacetylene           & 5.33       & 4.765       & 4.765      & 4.712      \\
\hline
 5   & Furan                 & 6.09       & 5.836       & 5.836      & 5.831      \\
\hline
 6   & Hexatriene            & 5.37       & 5.358       & 5.358      & 4.854      \\
\hline
 7   & Imidazole             & 5.71       & 5.651       & 5.651      & 5.458      \\
\hline
 8   & Pyrrole               & 5.24       & 5.170       & 5.170      & 5.052      \\
\hline
 9   & Thiophene             & 5.64       & 5.567       & 5.594      & 5.653      \\
\hline
     & AAE                   &            & 0.215       & 0.212      & 0.365      \\
\hline
\end{tabular}
\caption{Vertical excitation energies of the lowest SINGLET states (eV)
for 9 molecules with $g_{11}^{(k)} \approx g_{22}^{(k)}\approx 0$.}
\label{Table_sg}
\end{table}

In order to verify these results even further,
we have carried out additional simulations for
the best three molecules.  
Namely, we have performed very accurate simulations with multireference averaged quadratic 
coupled cluster (MR-AQCC) method \cite{MRAQCC}, as implemented in {\bf ORCA} \cite{Orca},
with def2-TZVPD basis set. 
The molecular geometries have been optimized using {\bf ORCA} tight optimization settings
with RI-mPW2PLYP/def2-TZVP (def2-TZVP/C) method. 
Afterwards, we have repeated the simulations within the mixed model of Sec. \ref{mixed} 
using the same geometries and the same basis set and with
the number of qubits in the environment $N_q$ varying in the range from 4 to 126.
The lowest eigenvalues of the Hamiltonian matrix at fixed $N_q$, having the dimensions $2^{N_q+2}\times 2^{N_q+2}$, 
have been computed with the Quantum Solver using the same approximation as we did earlier for diradicals. 
The results are presented in Fig. \ref{dETS} and in 
Table \ref{table222}, where we provide the vertical excitation energies of the lowest triplet states, 
$E_{\rm T}-E_{\rm S}$, obtained with 62 and with 126 qubits in the environment.
For $4\leq N_q\leq 62$ we have allowed up to 4 spin
excitations in the system (blue dots in Fig. \ref{dETS}), while for $63\leq N_q \leq 126$ we have allowed only up to 2 spin excitations
in order to save memory (orange dots in Fig. \ref{dETS}).  
We estimate the errors coming from reducing the size of the Hamiltonian matrix
as $10^{-5}$ eV for the blue dots and $10^{-3}$ eV for the orange dots, 
i.e. such errors are negligible.  According to Table \ref{table222}, 
the total error of our model simulations with both 62 and 126 qubits are well within the chemical accuracy of 1 kcal/mol (0.043 eV).
It is in agreement with the simulations based on the QUEST molecular geometries presented in Table \ref{Table_tr},
which further supports the validity of our findings.
Fig. \ref{dETS} shows the convergence of the triplet-singlet gaps with the 
growing number of the bath qubits $N_q$.  
This convergence turns out to be slow and even 126 qubits are insufficient to achieve it fully.
The dependence of the triplet-singlet gap on $N_q$ for thiophene exhibits a jump around $N_q=70$.
It is associated with a single bath qubit, among the 126 strongest coupled ones,  
with non-zero longitudinal coupling,  $g_{11}-g_{22}\not= 0$.

Thus, our model rather accurately describes selected molecules
with transverse coupling, i.e. those with $g_{11} \approx g_{22}\approx 0$ and $g_{12}\not=0$.
Vertical excitations for a number of molecules with non-vanishing longitudinal coupling, $g_{11}, g_{22}\not= 0$, are captured with lower
accuracy of $\sim 0.3$ eV. Further improvement of the model would require better treatment of
the interaction term $V_3$ (\ref{V3}) and introduction of new interaction
terms between the bath oscillators proportional to the Coulomb integrals
$h_{mm n\beta}$ and $h_{\alpha\alpha n\beta}$.

In conclusion, we have developed an approximate system-bath model suitable for quantum chemistry simulations.
It is particularly convenient for quantum computing algorithms because the 
Hamiltonian of a molecule is expressed in terms of qubit operators.
The "system" consists of the two HOMO/LUMO orbitals populated by two electrons, it is encoded
by two qubits. The "bath" consists of large number of either oscillators or qubits, with each
of them corresponding to an electron excitation from a doubly occupied orbital below the Fermi level to an empty orbital above it.
Oscillator and qubit baths are numerically equivalent due to very weak coupling between 
them and the system. We have demonstrated that with this model 
one can achieve chemical accuracy of 1 kcal/mol for the lowest vertical excitation
energies of selected molecules.

\section{Acknowledgment}

This work has received funding from the German Federal Ministry of Research, Technology and Space (BMFTR) within the PhoQuant project (Grant No. 13N16103) and the PASQUOPS project (Grant No. 13N17251).

\appendix

\section{Qubit representation of System Hamiltonian}
\label{System}

In this section we express the system Hamiltonian (\ref{Hsys})
in terms of qubit operators and bring it to the form (\ref{H_sys_qubits}).
We we also provide exact expressions for the eigen-energies and the eigenvectors of this Hamiltonian
in the limit $\tilde t_1=\tilde t_2=0$.

The Hilbert space of the system Hamiltonian (\ref{Hsys}) is fully determined by six states
\begin{eqnarray}
&& | \uparrow\downarrow, 0\rangle = c_{1\uparrow}^\dagger c_{1\downarrow}^\dagger |0\rangle, \;\;\; 
| 0 , \uparrow\downarrow \rangle = c_{2\uparrow}^\dagger c_{2\downarrow}^\dagger |0\rangle, \;\;\;
\nonumber\\ &&
| \uparrow, \downarrow\rangle = c_{1\uparrow}^\dagger c_{2\downarrow}^\dagger |0\rangle, \;\;\; 
| \downarrow, \uparrow \rangle = c_{1\downarrow}^\dagger c_{2\uparrow}^\dagger |0\rangle, \;\;\;
\nonumber\\ &&
| \uparrow, \uparrow \rangle = c_{1\uparrow}^\dagger c_{2\uparrow}^\dagger |0\rangle, \;\;\; 
| \downarrow, \downarrow \rangle = c_{1\downarrow}^\dagger c_{2\downarrow}^\dagger |0\rangle.
\label{states}
\end{eqnarray}
In this basis the Hamiltonian takes the form
\begin{widetext}
\begin{eqnarray}
H_{\rm sys}=\left(\begin{array}{cccc|cc}
 2 (\varepsilon_1+U_1) & K_{12}                & \tilde t_1                             & -\tilde t_1                            & 0                                             & 0 \\
 K_{12}                & 2 (\varepsilon_2+U_2) & \tilde t_2                             & -\tilde t_2                            & 0                                             & 0  \\
 \tilde t_1            & \tilde t_2            & \varepsilon_1 + \varepsilon_2 + J_{12} & - K_{12}                               & 0                                             & 0  \\
 -\tilde t_1           & -\tilde t_2           & -K_{12}                                & \varepsilon_1 + \varepsilon_2 + J_{12} & 0                                             & 0   \\
\hline
 0                     & 0                     & 0                                      & 0                                      & \varepsilon_1+\varepsilon_2 + J_{12} - K_{12} & 0    \\
 0                     & 0                     & 0                                      & 0                                      & 0                                             & \varepsilon_1 + \varepsilon_2 + J_{12} - K_{12}    \\
\end{array}\right).
\label{Hsys_6x6}
\end{eqnarray}
\end{widetext}

The Hamiltonian (\ref{Hsys_6x6}) is well known. For example, it describes
spin qubits based on semiconducting double quantum dots \cite{Burkhard1,Burkhard}
and provides the minimal model of a hydrogen molecule within the Hund-Mulliken molecular orbital theory.
The $6\times 6$ matrix (\ref{Hsys_6x6}) consists of a non-trivial $4\times 4$ sub-block in the upper left corner together with
two diagonal matrix elements in the lower right corner. The latter correspond to the energies of the two
fully spin polarized states $| \uparrow, \uparrow \rangle$ and $| \downarrow, \downarrow \rangle$, which 
represent the two components of the triplet state with the projections of the total spin equal, respectively, to $S_z=+1$ and $S_z=-1$.
The $4\times 4$ block in the upper left corner, which corresponds to the $S_z=0$ subspace, encodes the three singlet states together with 
the third component of the triplet state having the form $( | \uparrow, \downarrow\rangle + | \downarrow, \uparrow \rangle  )/\sqrt{2}$.
This $4\times 4$ block can be expressed in terms of the direct products of Pauli matrices representing two qubits
in the form (\ref{H_sys_qubits},\ref{H_sys_qubits_2}).

The matrix (\ref{Hsys_6x6}) can be diagonalized analytically \cite{RPA}, but the corresponding expressions are
rather lengthy. At the same time, for the majority of molecules one can put $\tilde t_1=\tilde t_2=0$. 
Indeed, $\tilde t_1$ and $\tilde t_2$ are the off-diagonal elements of the Fock matrix 
for the Hartree-Fock simulation of a closed shell singlet state. 
Even though we have initially run the restricted open-shell Hartree-Fock simulation for a triplet state,  
the amplitudes $\tilde t_1,\tilde t_2$ usually anyway remain very small.
Assuming $\tilde t_1=\tilde t_2=0$ one can easily find the eigen-values and the eigen-functions of the Hamiltonian (\ref{Hsys_6x6}).
Three of the eigenfunctions, namely,
\begin{eqnarray}
\Psi_1^{\rm tr} = | \uparrow,\uparrow \rangle, \;\; \Psi_0^{\rm tr} = \frac{| \uparrow,\downarrow \rangle + | \downarrow,\uparrow \rangle}{\sqrt{2}}, 
\;\; \Psi_{-1}^{\rm tr} = | \downarrow,\downarrow \rangle
\label{tr}
\end{eqnarray}
form the triplet state with the energy
\begin{eqnarray}
E^{\rm tr} = \varepsilon_1 + \varepsilon_2 + J_{12} - K_{12}.
\end{eqnarray}
The remaining three states are the singlets with the following eigenfunctions and energies:
\begin{eqnarray}
&& \Psi_1^{\rm sg} = \cos\theta_{12} | \uparrow\downarrow, 0 \rangle - \sin\theta_{12} | 0, \uparrow\downarrow \rangle, 
\nonumber\\ &&
E_1^{\rm sg} = \varepsilon_1 + \varepsilon_2 + U_1 + U_2 - \sqrt{\Delta_{12}^2 + K_{12}^2},
\label{sg1}
\end{eqnarray}
\begin{eqnarray}
&& \Psi_2^{\rm sg} = \frac{| \uparrow,\downarrow \rangle - | \downarrow,\uparrow \rangle}{\sqrt{2}}, 
\nonumber\\ &&
E_2^{\rm sg} = \varepsilon_1+\varepsilon_2 + J_{12} + K_{12}, 
\label{sg2}
\end{eqnarray}
\begin{eqnarray}
&& \Psi_3^{\rm sg} = \sin\theta_{12} | \uparrow\downarrow, 0 \rangle + \cos\theta_{12} | 0, \uparrow\downarrow \rangle, 
\nonumber\\ && 
E_3^{\rm sg} = \varepsilon_1 + \varepsilon_2 + U_1 + U_2 + \sqrt{\Delta_{12}^2 + K_{12}^2}.
\label{sg3}
\end{eqnarray}
Here we have introduced the angle $\theta_{12}$ such that
\begin{eqnarray}
\cos\theta_{12} &=& \frac{ \sqrt{\Delta_{12}^2 + K_{12}^2} + \Delta_{12} }{\sqrt{ K_{12}^2 + \left( \sqrt{\Delta_{12}^2 + K_{12}^2} + \Delta_{12} \right)^2 }},
\nonumber\\
\sin\theta_{12} &=& \frac{ K_{12} }{\sqrt{ K_{12}^2 + \left( \sqrt{\Delta_{12}^2 + K_{12}^2} + \Delta_{12} \right)^2 }}.
\label{theta12}
\end{eqnarray}

\section{Hamiltonian of the environment, transitions between the environment orbitals}
\label{Sec_env}

In this Appendix we approximately transform the Hamiltonian of the environment (\ref{Henv}) 
bringing it to the form of a bath of oscillators or a bath of two-level systems (qubits).  

In Sec. \ref{Ham} we have split the Hamiltonian in the contribution of the two HOMO/LUMO orbitals
and the remaining environment orbitals. Here we simplify the environment Hamiltonian (\ref{Henv})
by splitting it even further in the same way.
Namely, omitting the less important terms, we approximate the Hamiltonian (\ref{Henv}) as
\begin{eqnarray}
H_{\rm env} \approx \sum_{m,\alpha} H_{m\alpha} + V_{\rm env}.
\label{Henvapp}
\end{eqnarray}
Here the Hamiltonians of the individual pairs of doubly occupied and empty orbitals are
\begin{eqnarray}
&& H_{m\alpha} = \sum_\sigma\sum_{p_1,p_2=m,\alpha} t_{p_1p_2,\alpha} c^\dagger_{p_1\sigma} c_{p_2\sigma} 
\nonumber\\ &&
+ \sum_{\sigma\sigma'} \sum_{p_1,p_2,p_3,p_4=m,\alpha} \frac{h_{p_1p_2p_3p_4}}{2} c^\dagger_{p_1\sigma} c^\dagger_{p_3\sigma'} c_{p_4\sigma'} c_{p_2\sigma}, 
\label{Hmalpha}
\end{eqnarray}
with the hopping amplitudes
\begin{eqnarray}
t_{p_1p_2,\alpha} &=& t_{p_1p_2} + \sum_{k=1}^{N_{\rm db}} ( 2h_{p_1p_2kk} - h_{p_1kp_2k} )
\nonumber\\ &&
-\, 2h_{p_1p_2\alpha\alpha} - h_{p_1\alpha p_2\alpha}
\end{eqnarray}
containing the Hartree-Fock corrections coming from all doubly occupied environment orbitals
except $\alpha$. The interaction term in Eq. (\ref{Henvapp}) has the form
\begin{eqnarray}
V_{\rm env} &=& 
\frac{1}{2} \sum_{\sigma\sigma'} \sum_{m\alpha\not= n\beta} h_{m\alpha n\beta}  
( c^\dagger_{m\sigma} c_{\alpha\sigma} + c^\dagger_{\alpha\sigma} c_{m\sigma} )
\nonumber\\ && \times\,
( c^\dagger_{n\sigma'} c_{\beta\sigma'} + c^\dagger_{\beta\sigma'} c_{n\sigma'} ),
\label{V11}
\end{eqnarray}
where the sum runs over all possible pairs of indexes $m,\alpha$ and $n,\beta$ with identical
pairs $m=n,\alpha=\beta$ being excluded. Indeed, such terms  are already
accounted for by the Hamiltonians of the individual pairs (\ref{Hmalpha}). 
The Hamiltonian (\ref{V11}) contains only the leading interaction terms from 
the exact Hamiltonian (\ref{Henv}). In particular, the Hamiltonian (\ref{V11})
does not contain the terms with the Coulomb integrals of the form $h_{mn\alpha\beta}, h_{m_1n_1 m_2n_2}$ and $h_{\alpha_1\beta_1 \alpha_2\beta_2}$.
Such terms describe the transitions between two empty states or two doubly occupied states and 
give zero contribution to the system energy in the second order of perturbation theory.
We have also omitted those terms of the exact Hamiltonian (\ref{Henv}), which contain
the Coulomb integrals of the form $h_{mm n\beta},h_{\alpha\alpha n\beta}$. Such terms
introduce non-linear coupling between the bath oscillators and can be important 
for some molecules.

As a first step, we consider the Hamiltonian $H_{m\alpha}$ (\ref{Hmalpha}) of a single pair of the environment orbitals -- a doubly occupied orbital $\alpha$
and an empty orbital $m$. The Hamiltonian $H_{m\alpha}$ is equivalent to that
of the HOMO/LUMO orbitals (\ref{Hsys}). Therefore,
we can use the expression presented in Appendix \ref{System} and in the main text, replacing
the index 1 by $\alpha$ and the index 2 by $m$ in the appropriate places. 
Since, according to our assumption, the orbital $\alpha$ 
is doubly occupied, the ground state of the pair of orbitals $m,\alpha$ is the singlet with the energy (see Eq. (\ref{sg1}))
\begin{eqnarray}
E_0^{m\alpha} &=& \varepsilon_\alpha - 2U_\alpha + \varepsilon_m - 2J_{m\alpha} + K_{m\alpha}
\nonumber\\ &&
+\, U_\alpha + U_m - \sqrt{\Delta_{m\alpha}^2 + K_{m\alpha}^2}.
\label{E0ma}
\end{eqnarray}
Here the energy differences $\Delta_{m\alpha}$ are defined in Eq. (\ref{Dmalpha}),
and the energies of the orbital are 
\begin{eqnarray}
\varepsilon_\alpha &=& \tilde t_{\alpha\alpha} = t_{\alpha\alpha} + \sum_{k=1}^{N_{\rm db}} (2h_{\alpha\alpha kk} - h_{\alpha k \alpha k}),
\nonumber\\
\varepsilon_m &=& \tilde t_{mm} = t_{mm} + \sum_{k=1}^{N_{\rm db}} (2h_{mm kk} - h_{m k m k}).
\label{energies}
\end{eqnarray}
The summation here runs over $N_{\rm db}$ doubly occupied environment orbitals, the HOMO orbital is not included in the sum.
Comparing Eq. (\ref{energies}) and Eq. (\ref{sg1}) we note that the energies are replaced as
\begin{eqnarray}
\varepsilon_1 &\to& \varepsilon_\alpha - 2U_\alpha
\nonumber\\
\varepsilon_2 &\to & \varepsilon_m - 2J_{m\alpha} + K_{m\alpha}.
\label{shift}
\end{eqnarray}
The additional shifts $- 2U_\alpha$ and $- 2J_{m\alpha} + K_{m\alpha}$ are required 
to exclude the contribution of the orbital $\alpha$ from the sums in Eq. (\ref{energies}).
Next, the first excited state of the Hamiltonian $H_{m\alpha}$ is the triplet state 
similar to the state (\ref{tr}). 
However, 
due to spin conservation this triplet state cannot be excited by acting with the
interaction Hamiltonian (\ref{V11}) on the ground state singlet.
Instead, this operation excites the second singlet state  with the energy (cf. Eq. (\ref{sg2}))
\begin{eqnarray}
E_{2,m\alpha}^{\rm sg} &=&  \varepsilon_\alpha - 2U_\alpha + \varepsilon_m - 2J_{m\alpha} + K_{m\alpha} 
\nonumber\\ &&
+\, J_{m\alpha} + K_{m\alpha}.
\end{eqnarray}
Assuming that the coupling between the pairs, described by the Hamiltonian (\ref{V11}), is weak 
we expect that the third singlet state (\ref{sg3}) of the pair $m,\alpha$ is not excited. Therefore, 
we can approximately replace the
Hamiltonian $H_{m\alpha}$ with that of a two level system having the splitting between the energy levels
\begin{eqnarray}
&& \hbar\omega_{m\alpha} = E_{2,m\alpha}^{\rm sg} - E_0^{m\alpha} 
\nonumber\\ &&
= \sqrt{\Delta_{m\alpha}^2 + K_{m\alpha}^2} - U_\alpha - U_m + J_{m\alpha} + K_{m\alpha}.
\label{omega0}
\end{eqnarray}
The ground state energy $E_0^{m\alpha}$ (\ref{E0ma}) can be further split in the two parts: the Hartree-Fock energy $E_{\rm HF}^{m\alpha}$ and the correlation
energy $E_{\rm corr}^{m\alpha}$,
\begin{eqnarray}
E_0^{m\alpha} &=& E_{\rm HF}^{m\alpha} + E_{\rm corr}^{m\alpha},
\nonumber\\
E_{\rm HF}^{m\alpha} &=& 2\varepsilon_\alpha + 2U_\alpha,
\nonumber\\
E_{\rm corr}^{m\alpha} &=& \Delta_{m\alpha} - \sqrt{\Delta_{m\alpha}^2 + K_{m\alpha}^2}.
\end{eqnarray}
Thus, the Hamiltonian of the environment qubit describing the pair of orbitals $m,\alpha$ can be expressed in the form
\begin{eqnarray}
H^{m\alpha}_{\rm qb} = E_{\rm HF}^{m\alpha} + E_{\rm corr}^{m\alpha} + \hbar\omega_{m\alpha} \sigma_{m\alpha}^+ \sigma_{m\alpha}^-.
\label{Hqubit}
\end{eqnarray} 
Here $\sigma^\pm_{m\alpha} = ( \sigma^x_{m\alpha} \pm i \sigma^y_{m\alpha} )/2$ are the raising and lowering operators for the two level system, and
$\sigma^x_{m\alpha},\sigma^y_{m\alpha}$ are the Pauli matrices.
Since the third singlet state (\ref{sg3}) is not excited due to the weak coupling, with the same accuracy
we can treat the pair of orbitals as an oscillator and replace the qubit Hamiltonian (\ref{Hqubit}) by the oscillator one,
\begin{eqnarray}
H^{m\alpha}_{\rm osc} = E_{\rm HF}^{m\alpha} + E_{\rm corr}^{m\alpha} + \hbar\omega_{m\alpha} a_{m\alpha}^\dagger  a_{m\alpha}.
\label{HRPA}
\end{eqnarray}  
Here $a_{m\alpha},a^\dagger_{m\alpha}$ are the ladder operators of the oscillator corresponding to the transition 
between the orbitals $\alpha\to m$. Indeed, weak coupling also ensures that the second excited state of the oscillator
(\ref{HRPA}) is weakly excited. Hence the ground state and the first excited state of the oscillator 
provide sufficiently accurate description of its dynamics,
and the Hamiltonians (\ref{Hqubit}) and (\ref{HRPA}) become approximately equivalent. 
Modeling the transitions $\alpha\to m$ by oscillators 
corresponds to the random phase approximation (RPA) \cite{Bohm1,Bohm2,Bohm3,Furche}. 

Next, we consider the interaction between different pairs of orbitals.
We define the operator
\begin{eqnarray}
C^\dagger_{m\alpha} = \sum_\sigma c^\dagger_{m\sigma} c_{\alpha\sigma}.
\label{C}
\end{eqnarray}
Acting by this operator on the eigen-functions (\ref{tr},\ref{sg1},\ref{sg2},\ref{sg3}), in which the indexes are replaced
as $1\to \alpha$, $2\to m$, we obtain
\begin{eqnarray}
&& C^\dagger_{m\alpha}  | \uparrow\downarrow, 0 \rangle = \sqrt{2} \Psi_2^{\rm sg},\;\;
\nonumber\\ &&
C^\dagger_{m\alpha} \Psi_2^{\rm sg} = \sqrt{2} ( -\sin\theta_{m\alpha}\Psi_1^{\rm sg} + \cos\theta_{m\alpha}\Psi_3^{\rm sg} ),
\nonumber\\ &&
C^\dagger_{m\alpha} | 0,\uparrow\downarrow \rangle = 0.
\nonumber
\end{eqnarray}
Here the angle $\theta_{m\alpha}$ is defined in the same way as $\theta_{12}$, see Eq. (\ref{theta12}).
Hence, in the subspace of the two singlet wavefunctions $\Psi_1^{\rm sg}$ and $\Psi_2^{\rm sg}$, 
which have been used to define the Hamiltonian of the isolated pair (\ref{Hqubit}), the operator (\ref{C}) can be expressed in the matrix form 
\begin{eqnarray}
C^\dagger_{m\alpha} = \left( \begin{array}{cc} 0 & \sqrt{2}\cos\theta_{m\alpha}  \\  -\sqrt{2}\sin\theta_{m\alpha} & 0   \end{array}  \right). 
\end{eqnarray} 
This matrix can be further expressed via the qubit operators of the pair of orbitals as
\begin{eqnarray}
C^\dagger_{m\alpha} = \sqrt{2} \cos\theta_{m\alpha} \sigma^+_{m\alpha} - \sqrt{2} \sin\theta_{m\alpha} \sigma^-_{m\alpha}.
\label{C_q}
\end{eqnarray}
Hence, the interaction term (\ref{V11}) can be expressed as
\begin{eqnarray}
V_{\rm env} &=&  \sum_{m_1\alpha_1\not = m_2\alpha_2} r_{m_1\alpha_1}  h_{m_1\alpha_1 m_2\alpha_2} r_{m_2\alpha_2} 
\nonumber\\ && \times\,
( \sigma^+_{m_1\alpha_1} + \sigma_{m_1\alpha_1} )( \sigma^+_{m_2\alpha_2} + \sigma_{m_2\alpha_2} ),
\end{eqnarray}
where the factors $r_{m\alpha}$ are defined as
\begin{eqnarray}
r_{m\alpha} &=& \cos\theta_{m\alpha} - \sin\theta_{m\alpha}
\nonumber\\
&=& \frac{ \sqrt{\Delta_{m\alpha}^2 + K_{m\alpha}^2} + \Delta_{m\alpha} - K_{m\alpha}}
{\sqrt{ K_{m\alpha}^2 + \left( \sqrt{\Delta_{m\alpha}^2 + K_{m\alpha}^2} + \Delta_{m\alpha} \right)^2 }}.
\label{r}
\end{eqnarray} 
Combining all terms together, we obtain the Hamiltonian of the environment in the form
\begin{eqnarray}
H_{\rm env} &=&  E^{\rm env}_{\rm HF}	 
+  \sum_{m\alpha} \left(  E^{m\alpha}_{\rm corr} + \hbar\omega_{m\alpha} \sigma^+_{m\alpha} \sigma^-_{m\alpha}   \right)
\nonumber\\ &&
+\, \sum_{m\alpha\not = n\beta}  r_{m\alpha}  h_{m\alpha n\beta} r_{n\beta}\; \sigma^x_{m\alpha}  \sigma^x_{n\beta}.
\label{H_env_qubit}
\end{eqnarray}

If one models the environment as a bath of oscillators, then the same Hamiltonian becomes
\begin{eqnarray}
&& H_{\rm env} =  E^{\rm env}_{\rm HF}	 
+  \sum_{m\alpha} \left(  E^{m\alpha}_{\rm corr} + \hbar\omega_{m\alpha} a^\dagger_{m\alpha} a_{m\alpha}   \right)
\nonumber\\ &&
+\, \sum_{m\alpha\not = n\beta}  r_{m\alpha}  h_{m\alpha n\beta} r_{n\beta} 
( a^\dagger_{m\alpha} + a_{m\alpha} )( a^\dagger_{n\beta} + a_{n\beta} ).
\nonumber\\
\label{H_env_osc}
\end{eqnarray}
Note, that the environment oscillators interact with each other. In Appendix \ref{normal}
we will introduce the normal modes of the environment, which do not interact.

\section{Hamiltonian of a molecule with the singlet ground state of $H_{\rm sys}$}

In this section we assume that the system Hamiltonian (\ref{Hsys_6x6}) has the singlet ground state (\ref{sg1}).
Based on this assumption, we derive the approximate full Hamiltonian of the system and bring it to the 
form (\ref{H_qubit_env},\ref{H_sys_qubits_2},\ref{H_env_qubits},\ref{V_int_qubits}). 
The idea behind the approximations made in this section is simple: we treat the HOMO
orbital as one more doubly occupied orbital and the LUMO orbital -- as one more empty orbital
while considering the transitions between the HOMO/LUMO and the environment orbitals.  

\subsection{Approximating interaction between the HOMO/LUMO and the environment orbitals}

We begin with the terms $V_{\rm ex}$ (\ref{V_ex}) and $V_{\rm pair}$ (\ref{V_pair}). Their sum can be written in the form
\begin{eqnarray}
&& V_{\rm ex} + V_{\rm pair} = 
- \frac{1}{2}\sum_\sigma\sum_{p_1p_2=1}^{2} \sum_{k=1}^{N_{\rm env}} h_{p_1kp_2k} c^\dagger_{p_1\sigma} c_{p_2\sigma}
\nonumber\\ &&
-\, \frac{1}{2} \sum_\sigma \sum_{p=1}^{2} \sum_{k_1k_2=1}^{N_{\rm env}} h_{pk_1pk_2} c^\dagger_{k_1\sigma} c_{k_2\sigma}
\nonumber\\ &&
+\,\frac{1}{2}\sum_{\sigma_1\sigma_2}\sum_{p_1p_2=1}^{N_0} \sum_{k_1k_2=1}^{N_{\rm env}} 
h_{p_1k_1 p_2k_2} (c^\dagger_{p_1\sigma_1} c_{k_1\sigma_1} + c^\dagger_{k_1\sigma_1}c_{p_1\sigma_1})
\nonumber\\ && \times\,
(c^\dagger_{p_2\sigma_2} c_{k_2\sigma_2} + c^\dagger_{k_2\sigma_2}c_{p_2\sigma_2}). 
\end{eqnarray}
Averaging this expression over the environment operators, we obtain
\begin{eqnarray}
\langle V_{\rm ex} + V_{\rm pair} \rangle_{\rm env} = 
-\sum_{\sigma}\sum_{p_1p_2=1}^{2} \sum_{k=1}^{N_{\rm db}} 
h_{p_1k p_2k}   c^\dagger_{p_1\sigma} c_{p_2\sigma}.
\nonumber
\end{eqnarray}
This diagonal term is already included into the Hartree-Fock energies $\varepsilon_1,\varepsilon_2$ (\ref{E1E2},\ref{tilde_t}).
Then we can approximate
\begin{eqnarray}
&& V_{\rm ex} + V_{\rm pair} - \langle V_{\rm ex} + V_{\rm pair} \rangle_{\rm env} 
\nonumber\\ && = \,
\sum_{\sigma_1\sigma_2}\sum_{p_1p_2=1}^{2} \sum_{k_1k_2=1}^{N_{\rm env}} 
\frac{h_{p_1k_1 p_2k_2}}{2} (c^\dagger_{p_1\sigma_1} c_{k_1\sigma_1} + c^\dagger_{k_1\sigma_1}c_{p_1\sigma_1})
\nonumber\\ && \times\,
(c^\dagger_{p_2\sigma_2} c_{k_2\sigma_2} + c^\dagger_{k_2\sigma_2}c_{p_2\sigma_2}). 
\label{Vex_Vpair_0}
\end{eqnarray}

\begin{figure}
\includegraphics[width=0.9\columnwidth]{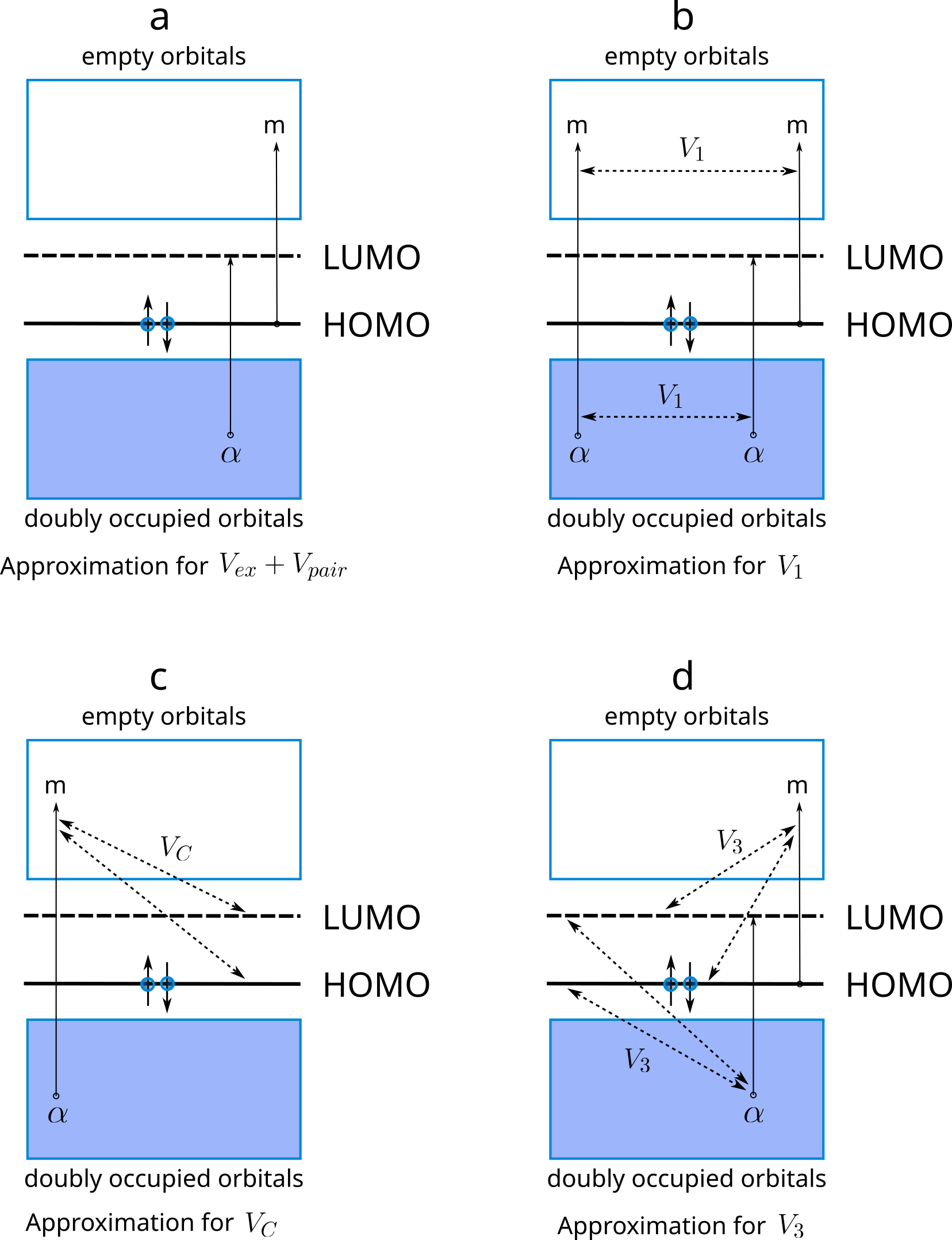}
\caption{Modeling of the interaction terms (\ref{Vint}).
(a) The terms $V_{\rm ex}+V_{\rm pair}$, defined in Eqs. (\ref{V_ex},\ref{V_pair}), are replaced 
by additional oscillators in the environment, indicated by arrows, which correspond to the 
transitions between HOMO/LUMO and environment orbitals. 
(b) The term $V_1$ (\ref{V1}) leads to interaction between
the oscillators describing the transitions between the environment orbitals (left arrow) and the oscillators
originating from transitions between HOMO/LUMO and environment orbitals. Dashed lines with arrows indicate this
interaction. 
(c) The Coulomb interaction term $V_{\rm C}$ (\ref{V_C}) leads to the interaction between the HOMO/LUMO orbitals
and the transitions involving the environment orbitals.
(d) The term $V_3$ (\ref{V3}) describes the interaction between the HOMO/LUMO orbitals 
and the oscillators coming from the terms $V_{\rm ex} + V_{\rm pair}$.}
\label{Fig_1}
\end{figure}

This expression looks very similar to the interaction term between the environment orbitals (\ref{V11}). However,
now two of the indexes, $p_1,p_2$, point to the HOMO and the LUMO orbitals. 
Here we assume that the ground state of the system Hamiltonian $H_{\rm sys}$ is the singlet state $\Psi_1^{\rm sg}$. 
Then, as in the previous section, we can approximately express the excitation operator  
in terms of the ladder operators of the oscillators,
\begin{eqnarray}
&& \sum_{\sigma_1} (c^\dagger_{1\sigma_1} c_{k_1\sigma_1} + c^\dagger_{k_1\sigma_1}c_{1\sigma_1})
= \sum_{\sigma_1} (c^\dagger_{1\sigma_1} c_{m\sigma_1} + c^\dagger_{m\sigma_1}c_{1\sigma_1})
\nonumber\\ &&
\to\, \sqrt{2} r_{m1} ( a^\dagger_{m1} + a_{m1}  ).
\end{eqnarray}
Here we keep only the transitions between the doubly occupied orbital 1 (HOMO)
and the empty orbitals $m$, which are shown by arrow in the right part of Fig. \ref{Fig_1}a. 
Thus, here we treat the HOMO orbital with the index 1 as yet another doubly occupied orbital of the environment. 
In the same way, we consider the orbital 2 (LUMO) as yet another empty orbital and make the replacement 
\begin{eqnarray}
&& \sum_{\sigma_1} (c^\dagger_{2\sigma_1} c_{k_1\sigma_1} + c^\dagger_{k_1\sigma_1}c_{2\sigma_1})
= \sum_{\sigma_1} (c^\dagger_{2\sigma_1} c_{\alpha\sigma_1} + c^\dagger_{\alpha\sigma_1}c_{2\sigma_1})
\nonumber\\ &&
\to\, \sqrt{2} r_{2\alpha} ( a^\dagger_{2\alpha} + a_{2\alpha}  ),
\end{eqnarray}
where, as before, the index $\alpha$ denotes a doubly occupied orbital of the environment.
The corresponding transition, $\alpha\to 2$, is also shown in Fig. \ref{Fig_1}a.
Combining these terms, we obtain several interaction terms between the newly introduced 
environment oscillators as shown in Fig. \ref{Fig_1}a, 
\begin{eqnarray}
&& V_{\rm ex} + V_{\rm pair} - \langle V_{\rm ex} + V_{\rm pair} \rangle_{\rm env} 
\nonumber\\ && \approx \,
\sum_{m_1m_2} r_{1m_1} h_{1m_1 1m_2} r_{1m2} ( a^\dagger_{m_11} + a_{m_11}  )( a^\dagger_{m_21} + a_{m_21}  )  
\nonumber\\ &&
+\, \sum_{\alpha_1\alpha_2} r_{2\alpha_1} h_{2\alpha_1 2\alpha_2} r_{2\alpha_2} ( a^\dagger_{\alpha_1 2} + a_{\alpha_1 2}  )( a^\dagger_{2\alpha_2} + a_{2\alpha_2}  )
\nonumber\\ &&
+\, \sum_{m\alpha} r_{1m} h_{1m 2\alpha} r_{2\alpha} ( a^\dagger_{m 1} + a_{m 1}  )( a^\dagger_{2\alpha} + a_{2\alpha}  )
\nonumber\\ &&
+\, \sum_{m\alpha} r_{2\alpha}h_{2\alpha 1m} r_{1m} ( a^\dagger_{2\alpha} + a_{2\alpha}  )( a^\dagger_{m 1} + a_{m 1}  ).
\label{Vex_Vpair}
\end{eqnarray}
These terms should be added to the Hamiltonian (\ref{H_env_osc}), which describes the transitions only between 
the environment orbitals $\alpha\to m$.
We should also add the corresponding kinetic terms 
\begin{eqnarray}
H^{\rm osc}_{m1,2\alpha} &=&  \sum_m (  E_{\rm corr}^{m1} + \hbar\omega_{m1} a^\dagger_{m1} a_{m1} )
\nonumber\\ &&
+\, \sum_\alpha (E_{\rm corr}^{2\alpha} + \hbar\omega_{2\alpha} a^\dagger_{2\alpha} a_{2\alpha}).
\label{H_env_ex_pair}
\end{eqnarray}
The frequencies of the oscillators and the correlation energies in these terms are
\begin{eqnarray}
\hbar\omega_{m1} &=& J_{m1} - U_1 - U_m + K_{m1} + \sqrt{\Delta^2_{m1} + K^2_{m1}},
\label{omega_m1}\\
\Delta_{m1} &=& \varepsilon_m + U_m - \varepsilon_1  - U_1,
\label{Delta_m1}\\
E_{\rm corr}^{m1} &=& \Delta_{m1} - \sqrt{\Delta^2_{m1} + K^2_{m1}},
\end{eqnarray}
\begin{eqnarray}
\hbar\omega_{2\alpha} &=& J_{2\alpha} - U_\alpha - U_2 + K_{2\alpha} + \sqrt{\Delta^2_{2\alpha} + K^2_{2\alpha}},
\label{omega_2a}\\
\Delta_{2\alpha} &=& \varepsilon_2 - \varepsilon_\alpha + U_2  + U_\alpha - 2 J_{2\alpha} + K_{2\alpha}  ,
\label{Delta_2a}\\
E_{\rm corr}^{2\alpha} &=& \Delta_{2\alpha} - \sqrt{\Delta^2_{2\alpha} + K^2_{2\alpha}}.
\end{eqnarray}
They are derived in the same way as the frequencies $\hbar\omega_{m\alpha}$ (\ref{omega0}).
Thus, we have approximately transformed the interaction terms $V_{\rm ex} + V_{\rm pair}$
into a set of additional oscillators of the environment interacting with each other. 
The corresponding transitions are shown in Fig. \ref{Fig_1}a.

Next, we consider the interaction term
\begin{eqnarray}
&& V_1 = \sum_{\sigma\sigma'} \sum_{p=1}^{2} \sum_{k_1k_2k_3=1}^{N_{\rm env}}  h_{pk_3 k_1k_2} 
\nonumber\\ && \times\,
\left(  c^\dagger_{p\sigma} c^\dagger_{k_1\sigma'} c_{k_2\sigma'} c_{k_3\sigma} +  c^\dagger_{k_3\sigma} c^\dagger_{k_2\sigma'} c_{k_1\sigma'} c_{p\sigma} \right).
\end{eqnarray} 
The average value of this term contributes to the off-diagonal components of the Fock matrix. Since this matrix is diagonal
for Hartree-Fock orbitals, we omit this contribution.
Employing the symmetries of the electron repulsion integrals (\ref{eri}), we bring this term to the form
\begin{eqnarray}
&& V_1 - \langle V_1 \rangle_{\rm env} = \sum_{\sigma\sigma'} \sum_{p=1}^{N_0} \sum_{k_1>k_2=1}^{N_{\rm env}} \sum_{k_3=1}^{N_{\rm env}} h_{pk_3 k_1k_2} 
\nonumber\\ && \times\,
\left(  c^\dagger_{p\sigma} c_{k_3\sigma} +  c^\dagger_{k_3\sigma}  c_{p\sigma} \right)
\left( c^\dagger_{k_1\sigma'} c_{k_2\sigma'} + c^\dagger_{k_2\sigma'} c_{k_1\sigma'} \right).
\nonumber\\
\end{eqnarray} 
We again leave only the terms in which the orbital $k_1$ is empty and the orbital $k_2$ is doubly occupied.
Renaming the indexes as $k_1\to m$, $k_2\to\alpha$ and using the property (see the previous section)
\begin{eqnarray}
&& \sum_{\sigma'} \left( c^\dagger_{m\sigma'} c_{\alpha\sigma'} + c^\dagger_{\alpha\sigma'} c_{m\sigma'} \right)
= C^\dagger_{m\alpha} + C_{m\alpha}
\nonumber\\ &&
\approx\, \sqrt{2} r_{m\alpha} (a^\dagger_{m\alpha} + a_{m\alpha}),
\label{app}
\end{eqnarray}
we can write the interaction term in the form
\begin{eqnarray}
&& V_1 - \langle V_1 \rangle_{\rm env} = \sqrt{2}\sum_{\sigma} \sum_{p=1}^{2}  \sum_{k_3=1}^{N_{\rm env}} \sum_{m\alpha} h_{pk_3 m\alpha} r_{m\alpha}
\nonumber\\ && \times\,
\left(  c^\dagger_{p\sigma} c_{k_3\sigma} +  c^\dagger_{k_3\sigma}  c_{p\sigma} \right)
\left( a^\dagger_{m\alpha} + a_{m\alpha} \right).
\end{eqnarray} 
Next, we transform the sum
\begin{eqnarray}
&& \sum_{k_3=1}^{N_{\rm env}}  h_{pk_3 m\alpha}\left(  c^\dagger_{p\sigma} c_{k_3\sigma} +  c^\dagger_{k_3\sigma}  c_{p\sigma} \right) 
\nonumber\\ &&
=\, \sum_{\beta=1}^{N_{\rm db}} h_{p\beta m\alpha} \left(  c^\dagger_{p\sigma} c_{\beta\sigma} +  c^\dagger_{\beta\sigma}  c_{p\sigma} \right)
\nonumber\\ &&
+\, \sum_{n=N_{\rm db}+1}^{N_{\rm env}}  h_{pn m\alpha}\left(  c^\dagger_{p\sigma} c_{n\sigma} +  c^\dagger_{n\sigma}  c_{p\sigma} \right).
\end{eqnarray}
According to our approximation, the orbital 1 can be viewed as doubly occupied and the orbital 2 as empty while we construct 
the oscillators of the environment. Therefore, performing the same analysis as in the previous section, we can write 
\begin{eqnarray}
&& \sum_{k_3=1}^{N_{\rm env}} h_{1k_3 m\alpha} \left(  c^\dagger_{1\sigma} c_{k_3\sigma} +  c^\dagger_{k_3\sigma}  c_{1\sigma} \right) 
\nonumber\\ &&
\approx\, \sum_{n={\rm empty}} h_{1n m\alpha} \left(  c^\dagger_{1\sigma} c_{n\sigma} +  c^\dagger_{n\sigma}  c_{1\sigma} \right)
\nonumber\\ &&
\approx\, \sqrt{2} \sum_{n} r_{1n} h_{1n m\alpha} (  a^\dagger_{1n} + a_{1n}  ).
\end{eqnarray}
Likewise, we transform
\begin{eqnarray}
&& \sum_{k_3=1}^{N_{\rm env}}  h_{2k_3 m\alpha}\left(  c^\dagger_{2\sigma} c_{k_3\sigma} +  c^\dagger_{k_3\sigma}  c_{2\sigma} \right)
\nonumber\\ &&
\approx\, \sqrt{2} \sum_\beta  r_{2\beta} h_{2\beta m\alpha} (  a^\dagger_{2\beta} + a_{2\beta}  ).
\end{eqnarray}
With these approximations, the interaction term can be written as
\begin{eqnarray}
&& V_1 - \langle V_1 \rangle_{\rm env} 
\nonumber\\ &&
=\, 2  \sum_n \sum_{m\alpha} r_{n1} h_{n1 m\alpha} r_{m\alpha} \left( a^\dagger_{n1} + a_{n1} \right) \left( a^\dagger_{m\alpha} + a_{m\alpha} \right)
\nonumber\\ &&
+\, 2 \sum_\beta  \sum_{m\alpha} r_{2\beta} h_{2\beta m\alpha} r_{m\alpha} \left( a^\dagger_{2\beta} + a_{2\beta} \right) \left( a^\dagger_{m\alpha} + a_{m\alpha} \right).
\nonumber\\
\label{V1_env}
\end{eqnarray} 
Thus, the term $V_1$ (\ref{V1}) describes the interaction between the oscillators describing the transitions between the environment orbitals 
and the oscillators involving the HOMO/LUMO orbitals, see Fig. \ref{Fig_1}b.

Next, we consider the interaction term $V_C$ (\ref{V_C}).
Following the procedure outlined above, we only keep the terms in which either the orbital $k_1$ is empty
and the orbital $k_2$ is doubly occupied or vice versa. After that, $V_C$ takes the form
\begin{eqnarray}
V_{\rm C} &=& \sum_{\sigma,\sigma'}\sum_{p_1p_2=1}^{2} \sum_{m\alpha} h_{p_1p_2 m\alpha}  c^\dagger_{p_1\sigma} c_{p_2\sigma}  
\nonumber\\ && \times\,
( c^\dagger_{m\sigma'} c_{\alpha\sigma'} + c^\dagger_{\alpha\sigma'} c_{m\sigma'} ).  
\end{eqnarray}
Performing the approximation (\ref{app}), we obtain
\begin{eqnarray}
V_{\rm C} &=& \sqrt{2}\sum_{\sigma,\sigma'}\sum_{p_1p_2=1}^{2} \sum_{m\alpha} h_{p_1p_2 m\alpha} r_{m\alpha}    
\nonumber\\ && \times\,
( a^\dagger_{m\alpha} + a_{m\alpha} ) c^\dagger_{p_1\sigma} c_{p_2\sigma}.  
\label{V_C_singlet}
\end{eqnarray}
Thus, the Coulomb term $V_{\rm C}$ describes the interaction between the oscillators 
describing the transitions between the environment orbitals and the system, i.e. the HOMO/LUMO orbitals, see Fig. \ref{Fig_1}c.

Finally, we consider the term $V_3$ (\ref{V3}), which is responsible for the interaction between the HOMO-LUMO orbitals and
the transitions from these orbitals and the environment, see Fig. \ref{Fig_1}d.
We transform this term to the form
\begin{eqnarray}
V_3 &=& \sum_{\sigma\sigma'} \sum_{p_1p_2p_3=1}^{2} \sum_{k=1}^{N_{\rm env}}  h_{kp_3p_1p_2} 
\nonumber\\ && \times\,
\left( c^\dagger_{k\sigma'}  c_{p_3\sigma'} + c^\dagger_{p_3\sigma'} c_{k\sigma'}  \right) c^\dagger_{p_1\sigma} c_{p_2\sigma}.
\end{eqnarray} 
As before, we assume that for $p_3=1$ the index $k$ should refer to an empty orbital $m$, and
for $p_3=2$ the index $k$ should point to a doubly occupied orbital $\alpha$. Then, the interaction term becomes
\begin{eqnarray}
V_3 &=& \sqrt{2}\sum_{\sigma} \sum_{p_1p_2=1}^{2}
\nonumber\\ &&
\bigg[ \sum_{m} r_{m 1} h_{m 1 p_1p_2} 
\left( a^\dagger_{m1} + a_{m1}  \right) c^\dagger_{p_1\sigma} c_{p_2\sigma}
\nonumber\\ &&
+\,  \sum_{\alpha} r_{2\alpha} h_{2\alpha p_1p_2} 
\left( a^\dagger_{2\alpha} + a_{2\alpha}  \right) c^\dagger_{p_1\sigma} c_{p_2\sigma}
\bigg].
\label{V3_env}
\end{eqnarray}

\subsection{Summary: Hamiltonian of a molecule with the singlet ground state of $H_{\rm sys}$}
\label{Sec_H_singlet}

Let us now summarize the results of this section. We have shown that the Hamiltonian of a molecule with the singlet ground state 
of the system Hamiltonian (\ref{Hsys}) may be approximated as
\begin{eqnarray}
&& H = H_{\rm sys} + H_{\rm env}  + \tilde V.
\label{Hsinglet}
\end{eqnarray}
Here the Hamiltonian $H_{\rm sys}$ is defined in Eq. (\ref{Hsys});
the environment Hamiltonian $H_{\rm env}$ is the sum of the Hamiltonian
(\ref{H_env_osc}), which describes the transitions between the environment orbitals,
and the additional terms (\ref{Vex_Vpair}), (\ref{H_env_ex_pair}) and (\ref{V1_env}),
which describe the newly added oscillators associated with the transitions between the HOMO/LUMO orbitals and
the environment. This sum can be cast to the form similar to that of Eq. (\ref{H_env_osc}),
\begin{eqnarray}
&& H_{\rm env} = E^{\rm env}_{\rm HF}	 
+  \sum_{m\alpha} \left(  E^{m\alpha}_{\rm corr} + \hbar\omega_{m\alpha} a^\dagger_{m\alpha} a_{m\alpha}   \right)
\nonumber\\ &&
+\, \sum_{m\alpha\not = n\beta}  
r_{m\alpha}  h_{m\alpha n\beta} r_{n\beta} 
( a^\dagger_{m\alpha} + a_{m\alpha} )( a^\dagger_{n\beta} + a_{n\beta} ),
\nonumber\\
\label{Henv_singlet}
\end{eqnarray}
but with the indexes $m,n$, $\alpha,\beta$ varying in slightly wider intervals
than they did in the Hamiltonian (\ref{H_env_osc}). Namely, now
the indexes $\alpha,\beta$ span over all doubly occupied orbitals and the HOMO orbital,
while the indexes $m,n$ go through all empty environment orbitals and the LUMO orbital.
The frequencies $\omega_{m\alpha}$, which correspond to the transitions between the
doubly occupied orbitals and the empty ones, are given by Eqs. (\ref{omega}), and
the frequencies for the transitions from or to the HOMO/LUMO orbitals are defined in Eqs. (\ref{omega_m1},\ref{omega_2a}). 
The environment Hamiltonian (\ref{Henv_singlet}) describes the oscillator environment, however,
one can easily convert it to the qubit environment Hamiltonian by replacing $a^\dagger_{m\alpha}\to \sigma^+_{m\alpha}$ and
$a_{m\alpha}\to \sigma^-_{m\alpha}$.

Finally, the interaction term in the Hamiltonian (\ref{Hsinglet}) is the sum of the terms $V_{\rm C}$, $V_3$
and the counter term required for restoring the normal ordering of the fermionic operators of the HOMO/LUMO orbitals
after the oscillators degrees of freedom are "integrated out" within the path integral approach.
The origin of this term will be discussed in Sec. \ref{counter}.
Thus, the interaction Hamiltonian $\tilde V$ has the form similar to that presented in Refs. \cite{Scott1,RPA},
\begin{eqnarray}
\tilde V &=& 
\sqrt{2}\sum_{\sigma}\sum_{pq=1}^{2} \sum_{m\alpha} h_{pq m\alpha} r_{m\alpha}    
( a^\dagger_{m\alpha} + a_{m\alpha} ) c^\dagger_{p\sigma} c_{q\sigma}
\nonumber\\ &&
+\, \sum_{\sigma}\sum_{pq=1}^{2} \sum_{l=1}^2 \frac{h_{pllq} - \tilde h_{pllq}}{2}  c^\dagger_{p\sigma} c_{q\sigma}.
\label{Vint_singlet}
\end{eqnarray}
Here the second line represents the counter term mentioned above, 
it is expressed in terms of the screened Coulomb integrals $\tilde h_{pllq}$ defined in Eq. (\ref{h_screened_2}).
The indexes $m$ and $\alpha$ vary in the same wider intervals as in the environment Hamiltonian (\ref{Henv_singlet}),
i.e. the interval for the index $m$ includes LUMO and the interval for $\alpha$ includes HOMO.
The total number of oscillators in the environment of a singlet molecule is
\begin{eqnarray}
N_{\rm RPA} = (N_{\rm db} + 1)(N_{\rm emp} + 1) - 1.
\label{N_RPA_singlet}
\end{eqnarray}
Here the last contribution, $-1$, implies the removal of the transition  HOMO $\rightarrow$ LUMO 
from the set of the environment oscillators. Indeed, this transition is described by the system Hamiltonian (\ref{H_sys_qubits})
and treated more accurately.

\section{Hamiltonian of a molecule with the triplet ground state of $H_{\rm sys}$}

In this section we consider molecules with the triplet ground state of the 
system Hamiltonian $H_{\rm sys}$. 
We emphasize that
the ground state of the full molecular Hamiltonian
$H_{\rm sys}+H_{\rm env}+\tilde V$
may become a singlet if the coupling between the HOMO/LUMO and
the environment orbitals is sufficiently strong to change the symmetry of the ground state.

\subsection{Approximating interaction between HOMO/LUMO and the environment orbitals}

As in the previous section, we approximate the interaction terms $V_{\rm ex},V_{\rm pair},V_{1},V_3$ 
by introducing additional oscillators in the environment. The transitions
corresponding to the environment oscillators of the triplet molecule are show in the right 
panel of Fig. \ref{Fig_exc}. 

In this section we assume that the ground state of the Hamiltonian $H_{\rm sys}$ is the triplet state $\Psi_1^{\rm tr}$ (\ref{tr}).
Acting on this wave-function by the operator $\sum_\sigma c^\dagger_{k\sigma}c_{1\sigma}$, we obtain
\begin{eqnarray}
&& \sum_\sigma c^\dagger_{k\sigma}c_{1\sigma} \Psi_1^{\rm tr} = \sum_\sigma c^\dagger_{k\sigma}c_{1\sigma} c^\dagger_{1\uparrow} c^\dagger_{2\uparrow} |0\rangle
= c^\dagger_{k\uparrow} c^\dagger_{2\uparrow} |0\rangle.
\nonumber
\end{eqnarray}
The resulting wave-function $c^\dagger_{k\uparrow} c^\dagger_{2\uparrow} |0\rangle$ differs from zero only if the orbital $k$ is empty. 
Therefore, considering the $2\times 2$ sub-space defined by
the functions $\Psi_1^{\rm tr}$ and $c^\dagger_{m\uparrow} c^\dagger_{2\uparrow} |0\rangle$ we can approximate
\begin{eqnarray}
\sum_\sigma c^\dagger_{m\sigma}c_{1\sigma} &\approx& \sigma^+_{m1},
\nonumber\\
\sum_\sigma ( c^\dagger_{m\sigma}c_{1\sigma} + c^\dagger_{1\sigma}c_{m\sigma} )   &\approx& \sigma^+_{m1} + \sigma^-_{m1}.
\label{app1}
\end{eqnarray}
where, as before, the index $m$ refers to an empty state. 
The energy change, corresponding to the excitation $1\to m$, is
\begin{eqnarray}
\hbar\omega_{m1} &=& \varepsilon_m + \varepsilon_2 + J_{m2} - K_{m2} - ( \varepsilon_1 + \varepsilon_2 + J_{12} - K_{12} )
\nonumber\\
&=& \varepsilon_m - \varepsilon_1 + J_{m2} - K_{m2} - J_{12} + K_{12}.
\end{eqnarray}

Now we assume that the environment orbital is doubly occupied and has the index $\alpha$. In this case
\begin{eqnarray}
&& \sum_\sigma c^\dagger_{1\sigma}c_{\alpha\sigma} \Psi_1^{\rm tr} = \sum_\sigma c^\dagger_{1\sigma}c_{\alpha\sigma} c^\dagger_{1\uparrow} c^\dagger_{2\uparrow} c^\dagger_{\alpha\uparrow}c^\dagger_{\alpha\downarrow}|0\rangle
\nonumber\\ &&
=\, c^\dagger_{1\downarrow} c^\dagger_{1\uparrow} c^\dagger_{2\uparrow} c_{\alpha\downarrow} c^\dagger_{\alpha\uparrow}c^\dagger_{\alpha\downarrow}|0\rangle
= - c^\dagger_{1\downarrow} c^\dagger_{1\uparrow} c^\dagger_{2\uparrow} c^\dagger_{\alpha\uparrow} |0\rangle
\nonumber\\ &&
=\, c^\dagger_{1\uparrow} c^\dagger_{1\downarrow} c^\dagger_{2\uparrow} c^\dagger_{\alpha\uparrow} |0\rangle.
\end{eqnarray}
We have obtained a triplet state, in which
the orbtials 2 and $\alpha$ are singly occupied and the orbital 1 became doubly occupied. Hence, 
in the $2\times 2$ Hilbert space defined by the wave-functions $\Psi_1^{\rm tr}$ and 
$c^\dagger_{1\uparrow} c^\dagger_{1\downarrow} c^\dagger_{2\uparrow} c^\dagger_{\alpha\uparrow} |0\rangle$ 
we can approximate
\begin{eqnarray}
&& \sum_\sigma c^\dagger_{1\sigma}c_{\alpha\sigma}  \approx \sigma^+_{1\alpha},
\nonumber\\
&& \sum_\sigma (c^\dagger_{1\sigma}c_{\alpha\sigma} +  c^\dagger_{\alpha\sigma}c_{1\sigma}) \approx \sigma^+_{1\alpha} + \sigma^-_{1\alpha}.
\label{app2}
\end{eqnarray} 

To obtain the energy change corresponding to the single electron excitation $\alpha\to 1$, we find the 
energies of the initial state $\Psi_1^{\rm tr}$
\begin{eqnarray}
E_i^{\rm tr} = E_{\rm HF}^{\rm env} + \varepsilon_1 + \varepsilon_2 + J_{12} - K_{12},
\end{eqnarray} 
and the energy of the final state $c^\dagger_{1\uparrow} c^\dagger_{1\downarrow} c^\dagger_{2\uparrow} c^\dagger_{\alpha\uparrow} |0\rangle$
\begin{eqnarray}
E_f^{\rm tr} = E^{\rm HF}_{2\alpha} + \tilde\varepsilon_\alpha + \tilde\varepsilon_2 + J_{2\alpha} - K_{2\alpha}.
\label{Ef}
\end{eqnarray} 
Here $E_{\rm HF}^{\rm env}$ is given by Eq. (\ref{EHF}) and 
$E^{\rm HF}_{2\alpha}$ is the Hartree-Fock energy with the orbitals $\alpha$ and $2$ excluded and
the orbital 1 doubly occupied. This energy can be expressed in the form similar to Eq. (\ref{EHF}),
but with modified occupation numbers of the orbitals $\tilde n_k$,
\begin{eqnarray}
E_{2\alpha}^{\rm HF} = 2\sum_k t_{k}\tilde n_k + \sum_{kk'} ( 2h_{kkk'k'} - h_{kk'kk'} ) \tilde n_{k} \tilde n_{k'}.
\end{eqnarray} 
Here $\tilde n_k=1$ for all doubly occupied orbitals of the environment except $\alpha$, for which $\tilde n_\alpha=0$, and $\tilde n_1=1$;
for all other orbitals $\tilde n_k=0$. Next, the energies $\tilde\varepsilon_\alpha$ and $\tilde\varepsilon_2$
are defined in the same way as the energies $\varepsilon_\alpha,\varepsilon_2$ (\ref{tilde_t},\ref{E1E2}) with the
adjustments corresponding to the replacement of the occupation numbers $n_\alpha\to \tilde n_\alpha$,  
\begin{eqnarray}
\tilde\varepsilon_\alpha &=& \varepsilon_\alpha - 2U_\alpha + 2J_{1\alpha} - K_{1\alpha},
\nonumber\\
\tilde\varepsilon_2 &=& \varepsilon_2 - 2J_{2\alpha} + K_{2\alpha} + 2J_{12} - K_{12}.
\end{eqnarray}
Now we can find the excitation energy of the transition $\alpha\to 1$,
\begin{eqnarray}
\hbar\omega_{1\alpha} &=& E^{\rm HF}_{2\alpha} + \tilde\varepsilon_2 + \tilde\varepsilon_\alpha + J_{2\alpha} - K_{2\alpha}
\nonumber\\ &&
-\, E^{\rm HF}_{12} - ( \varepsilon_1 + \varepsilon_2 + J_{12} - K_{12} ).
\end{eqnarray}
Since the difference between the Hartree-Fock energies has the form
\begin{eqnarray}
&& E^{\rm HF}_{2\alpha} - E^{\rm HF}_{12}
\nonumber\\ && 
=\, 2 ( \varepsilon_1 - \varepsilon_\alpha )
+ 2U_1 + 2U_\alpha - 4J_{1\alpha} + 2K_{1\alpha} ,
\end{eqnarray}
we obtain
\begin{eqnarray}
\hbar\omega_{1\alpha} = 
\varepsilon_1 - \varepsilon_\alpha  + 2U_1 - 2J_{1\alpha} + K_{1\alpha} - J_{2\alpha}  + J_{12}.
\end{eqnarray} 
Here we summarize the excitation energies for all additional transitions depicted in
right panel of Fig. \ref{Fig_exc}:
\begin{eqnarray}
\hbar\omega_{m1} &=& \varepsilon_m - \varepsilon_1 + J_{m2} - K_{m2} - J_{12} + K_{12},
\nonumber\\
\hbar\omega_{1\alpha} &=& \varepsilon_1 - \varepsilon_\alpha  + 2U_1 - 2J_{1\alpha} + K_{1\alpha} - J_{2\alpha}  + J_{12},
\nonumber\\
\hbar\omega_{m2} &=& \varepsilon_m - \varepsilon_2 + J_{m1} - K_{m1} - J_{12} + K_{12},
\nonumber\\
\hbar\omega_{2\alpha} &=& \varepsilon_2 - \varepsilon_\alpha  + 2U_2 - 2J_{2\alpha} + K_{2\alpha} - J_{1\alpha}  + J_{12}.
\nonumber\\
\label{omega_ij}
\end{eqnarray}
The frequencies $\hbar\omega_{m2}$ and $\hbar\omega_{2\alpha}$ are found in the same way as $\hbar\omega_{m1}$ and $\hbar\omega_{1\alpha}$.
The frequencies of the transitions between the environment orbitals $\hbar\omega_{m\alpha}$ are given by the
same expression (\ref{omega},\ref{omega0}) as in the case of molecules with the singlet ground state of the Hamiltonian $H_{\rm sys}$
because they are determined by the same Hamiltonian $H_{\rm env}$ decoupled from $H_{\rm sys}$.

With these results at hand, we obtain the approximate expression for the interaction terms
\begin{eqnarray}
&& V_{\rm ex} + V_{\rm pair} - \langle V_{\rm ex} + V_{\rm pair} \rangle_{\rm env}  \approx \,
\frac{1}{2}\sum_{p_1p_2=1}^2
\nonumber\\ && \times\,
\bigg[\sum_{m_1m_2} h_{p_1 m_1 p_2 m_2} ( a^\dagger_{m_1 p_1} + a_{m_1 p_1}  )( a^\dagger_{m_2 p_2} + a_{m_2 p_2}  )  
\nonumber\\ &&
+\, \sum_{\alpha_1\alpha_2} h_{p_1 \alpha_1 p_2\alpha_2} ( a^\dagger_{p_1\alpha_1} + a_{p_1\alpha_1}  )( a^\dagger_{p_2\alpha_2} + a_{p_2\alpha_2}  )
\nonumber\\ &&
+\, \sum_{m\alpha} h_{p_1 m p_2\alpha} ( a^\dagger_{m p_1} + a_{m p_1}  )( a^\dagger_{p_2\alpha} + a_{p_2\alpha}  )
\nonumber\\ &&
+\, \sum_{m\alpha} h_{p_1\alpha p_2 m} ( a^\dagger_{p_1\alpha} + a_{p_1\alpha}  )( a^\dagger_{m p_2} + a_{m p_2}  )\bigg].
\label{Vex_Vpair_triplet}
\end{eqnarray}
Here we have repeated the analysis outlined above for the excitations $2\to m$ and $\alpha\to 2$
and have replaced the Pauli matrices $\sigma^\pm$, appearing in the expressions (\ref{app1},\ref{app2}), with the  ladder operators of the oscillators $a,a^\dagger$.
We note that the interaction term $V_{\rm ex} + V_{\rm pair}$ for the triplet ground state (\ref{Vex_Vpair_triplet})
contains more oscillators than the corresponding term for a singlet molecule (\ref{Vex_Vpair}), and that the pre-factors
in front of these terms are two times smaller.

Next, we consider the interaction term $V_1$ (\ref{V1}).
Repeating the analysis of the previous section with proper modifications, we arrive at 
the following approximation for $V_1$ in terms of the environment oscillators
\begin{eqnarray}
&& V_1 - \langle V_1 \rangle_{\rm env} = 
\nonumber\\ &&
=\, \sqrt{2} \sum_{p=1}^{2} \sum_{mn\alpha} h_{np m\alpha}  r_{m\alpha}
\left(  a^\dagger_{np} + a_{np} \right) \left( a^\dagger_{m\alpha} + a_{m\alpha} \right)
\nonumber\\ &&
+\, \sqrt{2} \sum_{p=1}^{2}  \sum_{m\alpha\beta} h_{p\beta m\alpha}  r_{m\alpha}
\left(  a^\dagger_{p\beta} + a_{p\beta} \right)\left( a^\dagger_{m\alpha} + a_{m\alpha} \right).
\nonumber\\
\end{eqnarray} 
This expression also differs from the corresponding formula for the singlet state (\ref{V1_env}) by a pre-factor.

The term $V_{\rm C}$, describing the interaction between the HOMO/LUMO orbitals and the oscillators corresponding to the transitions
between the environment orbitals,  retains the same form as in the case of singlet molecules (\ref{V_C_singlet}).

Finally, we consider the term $V_3$, which is responsible for the interaction between the HOMO/LUMO orbitals
and the oscillators of the environment arising from the transitions between HOMO/LUMO and all other orbitals.
Repeating the same steps as before, we obtain this term in the form   
\begin{eqnarray}
&& V_3 =  \sum_{\sigma} \sum_{p_1p_2p_3=1}^{2} \sum_{m}  h_{mp_3p_1p_2} \left( a^\dagger_{mp_3} + a_{mp_3}  \right) c^\dagger_{p_1\sigma} c_{p_2\sigma}
\nonumber\\ &&
+\,  \sum_{\sigma} \sum_{p_1p_2p_3=1}^{2} \sum_{\alpha}  h_{\alpha p_3p_1p_2} 
\left( a^\dagger_{p_3\alpha} + a_{p_3\alpha}  \right) c^\dagger_{p_1\sigma} c_{p_2\sigma}.
\nonumber\\
\label{V3_app}
\end{eqnarray} 

\subsection{Summary: Hamiltonian of a molecule with the triplet ground state of $H_{\rm sys}$}
\label{Sec_H_trilpet}

Summarizing the results of the previous subsection, here we present the full Hamiltonian of a molecule
with the triplet ground state of the system Hamiltonian  $H_{\rm sys}$.
It has the form 
\begin{eqnarray}
&& H = H_{\rm sys} + H_{\rm env}  + \tilde V,
\label{H_triplet}
\end{eqnarray}
with $H_{\rm sys}$ defined in Eq. (\ref{Hsys}),
the environment Hamiltonian having the form
\begin{eqnarray}
&& H_{\rm env} =  E^{\rm env}_{\rm HF}	 
+  \sum_{m\alpha} \left( E_{\rm corr}^{m\alpha} + \hbar\omega_{m\alpha} a^\dagger_{m\alpha} a_{m\alpha}  \right)
\nonumber\\ &&
+\, \sum_{m\alpha\not = n\beta}  
r_{m\alpha}  h_{m\alpha n\beta} r_{n\beta} 
( a^\dagger_{m\alpha} + a_{m\alpha} )( a^\dagger_{n\beta} + a_{n\beta} )
\nonumber\\ &&
+\, \sum_{p=1}^2 \left(\sum_m \hbar\omega_{mp} a^\dagger_{mp} a_{mp} + \sum_\alpha \hbar\omega_{p\alpha} a^\dagger_{p\alpha} a_{p\alpha}\right)
\nonumber\\ &&
+\, \frac{1}{2}\sum_{p,q=1}^2\sum_{mn} h_{p m q n} ( a^\dagger_{m p} + a_{m p}  )( a^\dagger_{n q} + a_{n q}  )  
\nonumber\\ &&
+\, \frac{1}{2}\sum_{p,q=1}^2\sum_{\alpha\beta} h_{p \alpha q\beta} ( a^\dagger_{p\alpha} + a_{p\alpha}  )( a^\dagger_{q\beta} + a_{q\beta}  )
\nonumber\\ &&
+\, \frac{1}{2}\sum_{p,q=1}^2\sum_{m\alpha} h_{p m q\alpha} ( a^\dagger_{m p} + a_{m p}  )( a^\dagger_{q\alpha} + a_{q\alpha}  )
\nonumber\\ &&
+\, \frac{1}{2}\sum_{p,q=1}^2\sum_{m\alpha} h_{p\alpha q m} ( a^\dagger_{p\alpha} + a_{p\alpha}  )( a^\dagger_{m q} + a_{m q}  )
\nonumber\\ &&
+\, \sqrt{2} \sum_{p=1}^{2} \sum_{mn\alpha} h_{np m\alpha}  r_{m\alpha}
\left(  a^\dagger_{np} + a_{np} \right) \left( a^\dagger_{m\alpha} + a_{m\alpha} \right)
\nonumber\\ &&
+\, \sqrt{2} \sum_{p=1}^{2}  \sum_{m\alpha\beta} h_{p\beta m\alpha}  r_{m\alpha}
\left(  a^\dagger_{p\beta} + a_{p\beta} \right)\left( a^\dagger_{m\alpha} + a_{m\alpha} \right),
\nonumber\\
\label{H_env_triplet}
\end{eqnarray}
and the interaction term
\begin{eqnarray}
\tilde V &=& 
\sum_{\sigma}\sum_{p_1p_2=1}^{2} \sum_{q=1}^2 \frac{h_{p_1qqp_2} - \tilde h_{p_1qqp_2}}{2}  c^\dagger_{p_1\sigma} c_{p_2\sigma}
\nonumber\\ &&
+\, \sum_{\sigma}\sum_{p,q=1}^{2} \bigg[
\sum_{m\alpha} \sqrt{2} h_{pq m\alpha} r_{m\alpha}    
( a^\dagger_{m\alpha} + a_{m\alpha} ) 
\nonumber\\ &&
+\, \sum_{l=1}^{2} \sum_{m} h_{pq ml} ( a^\dagger_{ml} + a_{ml} ) 
\nonumber\\ &&
+\, \sum_{l=1}^{2} \sum_{\alpha} h_{pq l\alpha} ( a^\dagger_{l\alpha} + a_{l\alpha} )  \bigg] c^\dagger_{p\sigma} c_{q\sigma}.
\label{Vint_triplet}
\end{eqnarray}
The first term of the interaction Hamiltonian $\tilde V$ is the counter-term having the same form as in the case of singlet molecules, see Eq. (\ref{Vint_singlet}).
In contrast to the Hamiltonian of singlet molecules, here the index $\alpha$ covers only the doubly occupied orbitals
and the index $m$ -- only the empty ones, i.e. the HOMO/LUMO orbitals are excluded from the sums. 
The frequencies $\omega_{m\alpha}$, which correspond the transitions between the
doubly occupied orbitals and the empty ones, are given by Eqs. (\ref{omega}), and
the frequencies for the transitions from or to the HOMO/LUMO orbitals are defined in Eqs. (\ref{omega_ij}).
The total number of the environment oscillators for a triplet molecule is
\begin{eqnarray}
N_{\rm RPA} = (N_{\rm db} +2)(N_{\rm empt}+2) - 4.
\label{N_RPA_triplet}
\end{eqnarray}

\section{Diagonalizing the environment}
\label{normal}

In this section we introduce the normal modes of the environment, which diagonalize the environment Hamiltonians (\ref{Henv_singlet},\ref{H_env_triplet}). 
For definiteness, we consider the environment Hamiltonian of a singlet molecule (\ref{Henv_singlet}), the triplet case is handled in the same way with
slight modifications.
We treat every pair of indexes $m_1,\alpha_1$ or $m_2,\alpha_2$ as a single index. 
We define the square matrix $\hat M$, which has the dimensions $N_{\rm RPA}\times N_{\rm RPA}$ and has the matrix elements (\ref{M}).
We also introduce the frequencies of the normal modes $\Omega_{m\alpha}$, which are found from the equation (\ref{Omega_ma}).
Next, we define the matrix $\hat S$, in which every column is an eigenvector of the matrix $\hat M$ and
the diagonal matrices of frequencies $\hat\Omega$ and $\hat\omega$, with the diagonal matrix elements $\Omega_{m\alpha}$
and $\omega_{m\alpha}$ respectively. The matrices $\hat M$ and $\hat S$ are related by the condition $\hat M\hat S = \hat S\hat\Omega^2$. 
Afterwards, we can express the ladder operators of the normal modes of the environment $b_{m\alpha},b^\dagger_{m\alpha}$
in terms of the ladder operators $a_{m\alpha},a^\dagger_{m\alpha}$, which describe the original uncoupled transitions in the environment, as follows 
\begin{eqnarray}
{\bm b} = \frac{\hat\Omega^{\frac{1}{2}} \hat S^T \hat\omega^{-\frac{1}{2}} ( {\bm a}^\dagger + {\bm a} ) - \hat\Omega^{-\frac{1}{2}} \hat S^T \hat\omega^{\frac{1}{2}} ( {\bm a}^\dagger - {\bm a} )}{2},
\nonumber\\
{\bm b}^\dagger = \frac{\hat\Omega^{\frac{1}{2}} \hat S^T \hat\omega^{-\frac{1}{2}} ( {\bm a}^\dagger + {\bm a} ) + \hat\Omega^{-\frac{1}{2}} \hat S^T \hat\omega^{\frac{1}{2}} ( {\bm a}^\dagger - {\bm a} )}{2}.
\label{bb+}
\end{eqnarray}
Here we have combined all operators $b_{m\alpha}$ and $a_{m\alpha}$ into the column vectors ${\bm b},{\bm a}$ with the dimension $N_{\rm RPA}$.
The inverse transformation for the operators reads
\begin{eqnarray}
{\bm a} = \frac{\hat\omega^{\frac{1}{2}} \hat S \hat\Omega^{-\frac{1}{2}} ( {\bm b}^\dagger + {\bm b} ) - \hat\omega^{-\frac{1}{2}}\hat S \hat\Omega^{\frac{1}{2}} ( {\bm b}^\dagger - {\bm b} )}{2},
\nonumber\\
{\bm a}^\dagger = \frac{\hat\omega^{\frac{1}{2}} \hat S \hat\Omega^{-\frac{1}{2}} ( {\bm b}^\dagger + {\bm b} ) + \hat\omega^{-\frac{1}{2}} \hat S \hat\Omega^{\frac{1}{2}} ( {\bm b}^\dagger - {\bm b} )}{2}.
\label{aa+}
\end{eqnarray}

After the transformation (\ref{aa+}) the full Hamiltonian takes the form
\begin{eqnarray}
H = H_{\rm sys} + H_{\rm env} + \tilde V,
\end{eqnarray}
where the Hamiltonian of the system is given by Eq. (\ref{Hsys}),
the Hamiltonian of the environment acquires the form (\ref{Henv_diag}),
and the interaction term (\ref{Vint_singlet}) becomes
\begin{eqnarray}
\tilde V &=& \sum_{\sigma}\sum_{p_1p_2=1}^{2} \sum_{n\beta} g_{p_1p_2}^{n\beta}    
( b^\dagger_{n\beta} + b_{n\beta} ) c^\dagger_{p_1\sigma} c_{p_2\sigma}
\nonumber\\ &&
+\,\sum_{\sigma}\sum_{p_1p_2=1}^{2} \sum_{q=1}^2 \sum_{n\beta} \frac{g_{p_1 q}^{n\beta} g_{qp_2}^{n\beta}}{\hbar\Omega_{n\beta}} c^\dagger_{p_1\sigma} c_{p_2\sigma}.
\label{Vint_normal_0}
\end{eqnarray}
Here the second line is a counter-term added to the Hamiltonian to ensure
normal ordering of the operators in the static screening limit (see Sec. \ref{counter}) 
and the coupling constants $g_{p_1p_2}^{n\beta}$ are given by Eq. (\ref{g}).
Expressions (\ref{Henv_diag}) and (\ref{Vint_normal_0}) also remain valid for the  molecules
with the triplet ground state of $H_{\rm sys}$.
In this case one should just modify the matrix $M$ (\ref{M}) and the coupling constants
(\ref{g}) in such a way that they agree with the environment Hamiltonian (\ref{H_env_triplet})
and with the interaction term (\ref{Vint_triplet}).

The interaction Hamiltonian (\ref{Vint_normal_0}) can be expressed in terms of qubit operators
encoding the HOMO/LUMO orbitals
if one writes it in the basis of the states (\ref{states}) and uses the same
rules of converting the resulting $4\times 4$ matrix as for the
system Hamiltonian (\ref{H_sys_qubits}). Afterwards, it acquires the form (\ref{V_int_normal})
for the oscillator environment or (\ref{V_int_qubits}) for the qubit environment.

\begin{table*}
\begin{tabular}{|l|l||c|c||c|c||c|}
\hline
     & Molecule              & lowest triplet & lowest triplet & lowest singlet & lowest singlet  & model of the \\ 
     &                       & QUEST          & static app.    & QUEST          & static app.     & environment  \\
\hline
 1   & Acetaldehyde          & 3.97           &  3.60         & 4.31           &  3.93          &  tr         \\
\hline
 2   & Acetone               & 4.13           &  3.77         & 4.47           &  4.08          &  tr         \\
\hline
 3   & Benzene               & 4.16           &  4.45         & 5.06           &  5.44          &  sg          \\
\hline
 4   & Butadiene             & 3.36           &  3.02         & 6.22           &  5.74          &  sg          \\
\hline
 5   & Cyanoformaldehyde     & 3.44           &  3.17         & 3.81           &  3.51          &  tr          \\
\hline
 6   & Cyclopentadiene       & 3.31           &  3.36         & 5.54           &  5.50          &  sg          \\
\hline
 7   & Diacetylene           & 4.10           &  4.33         & 5.33           &  4.77          &  sg          \\
\hline
 8   & Ethylene              & 4.54           &  5.18         & 7.39           &  7.17          &  sg          \\
\hline
 9   & Formaldehyde          & 3.58           &  3.31         & 3.98           &  3.65          &  tr          \\
\hline
 10  & Furan                 & 4.20           &  4.12         & 6.09           &  5.84          &  sg          \\
\hline
 11  & Hexatriene            & 2.73           &  2.58         & 5.37           &  5.36          &  sg          \\
\hline
 12  & Imidazole             & 4.73           &  4.89         & 5.71           &  5.65          &  sg          \\
\hline
 13  & Izobutene             & 4.53           &  5.25         & 6.46           &  6.91          &  sg          \\
\hline
 14  & Ketene                & 3.77           &  3.42         & 3.85           &  3.66          &  sg          \\
\hline
 15  & Metylenecyclopropene  & 3.49           &  3.85         & 4.28           &  4.43          &  sg          \\
\hline
 16  & Nitrosomethane        & 1.16           &  1.43         &  1.96          &  1.70          &  tr          \\
\hline
 17  & Nitroxyl (HNO)        & 0.88           &  0.85         &  1.74          &  1.52          &  tr          \\
\hline
 18  & Pyrrole               & 4.51           &  4.58         &  5.24          &  5.17          &  sg          \\
\hline
 19  & Thioformaldehyde      & 1.94           &  1.32         &  2.22          &  1.57          &  tr          \\
\hline
 20  & Thiophene             & 3.97           &  4.02         &  5.64          &  5.57          &  sg          \\
\hline
     & AAE$^*$               &  0             &  0.28         &   0            &  0.27          &              \\
\hline
\end{tabular}
\caption{Vertical excitation energies (eV) of the lowest triplet and of the lowest singlet states for a set of 20 molecules
computed with the static approximation of Sec. \ref{Sec_static}. 
The reference values for the energies and the geometries of the molecules are taken from the QUEST database \cite{QUEST}. 
Basis set def2-TZVPD has been used for the simulations. \\
$^*$ AAE stands for the absolute average error.}
\label{table2}
\end{table*}

\section{Static screening approximation and the counter term}
\label{counter}

The frequencies of the environment oscillators usually exceed the
energy splitting between the HOMO and the LUMO orbitals, $\hbar\Omega_{m\alpha} \gg \Delta_{12}$.
Therefore, one can use the static screening approximation introduced in Ref. \cite{RPA}. 
In this approximation, the effect of the environment  is absorbed
in the Lamb shift correction \cite{BP} to the system Hamiltonian as
\begin{eqnarray}
\tilde H_{\rm sys} = H_{\rm sys} + \langle \tilde V \rangle_{\rm env}
- \frac{i}{\hbar}\int_0^\infty dt \langle \delta \tilde V(0) \delta \tilde V(-t)  \rangle_{\rm env},
\label{Lamb}\\ 
\delta \tilde V = \tilde V -  \langle \tilde V \rangle_{\rm env}. \hspace{5.2cm}
\end{eqnarray} 
The second term in Eq. (\ref{Lamb}) is the Lamb shift \cite{BP} and the average value $\langle \tilde V \rangle_{\rm env}$
is the counter-term, which is required to ensure the normal ordering of the fermionic operators in the exact system Hamiltonian. 
Writing down the Lamb shift for the interaction Hamiltonian (\ref{Vint_normal_0}), we obtain
\begin{eqnarray}
&& -\frac{i}{\hbar}\int_0^\infty dt \langle \delta \tilde V(0) \delta \tilde V(-t)  \rangle_{\rm env} = 
\nonumber\\ &&
=\,- \frac{i}{\hbar}\int_0^\infty dt
\sum_{\sigma\sigma'}\sum_{p_1p_2p_3p_4=1}^{2} \sum_{m\alpha,n\beta} g_{p_1p_2}^{m\alpha} g_{p_3p_4}^{n\beta}
\nonumber\\ &&\times\,    
c^\dagger_{p_1\sigma}(0) c_{p_2\sigma}(0) c^\dagger_{p_3\sigma'}(-t) c_{p_4\sigma'}(-t)
\nonumber\\ && \times\,
\langle ( b^\dagger_{m\alpha}(0) + b_{m\alpha}(0) ) ( b^\dagger_{n\beta}(-t) + b_{n\beta}(-t) ) \rangle.
\nonumber
\end{eqnarray} 
The static screening approximation relies on the assumption that the time integral in this expression converges
fast and one can make Markovian approximation replacing $c^\dagger_{p_3\sigma'}(-t) c_{p_4\sigma'}(-t) \to c^\dagger_{p_3\sigma'}(0) c_{p_4\sigma'}(0)$. 
As we mentioned above, this approximation is sufficient if $\hbar\Omega_{m\alpha} \gg \Delta_{12}$ for
all oscillators of the environment. After that, finding the Lamb shift reduces to the evaluation of the following
time integral:
\begin{eqnarray}
&& \int_0^\infty dt \langle ( b^\dagger_{m\alpha}(0) + b_{m\alpha}(0) ) ( b^\dagger_{n\beta}(-t) + b_{n\beta}(-t) ) \rangle 
\nonumber\\ &&
=\, \int_0^\infty dt \left\langle ( b^\dagger_{m\alpha} + b_{m\alpha} ) \left( e^{-i\Omega_{n\beta}t} b^\dagger_{n\beta} + e^{i\Omega_{n\beta}t} b_{n\beta} \right) \right\rangle
\nonumber\\ &&
=\,\int_0^\infty dt e^{-i\Omega_{m\alpha}t} \delta_{m\alpha,n\beta} = \frac{\delta_{m\alpha,n\beta}}{i\Omega_{m\alpha}}.
\label{corr}
\end{eqnarray} 
Here we have assumed that the environment oscillators are in their ground state.
We also emphasize that the same result holds for the qubit environment.

Combining these expressions, 
we obtain the renormalized Hamiltonian (\ref{Lamb}) in the form
\begin{eqnarray}
\tilde H_{\rm sys} = H_{\rm sys}
+\sum_{\sigma}\sum_{p_1p_2=1}^{2} \sum_{q=1}^2 \sum_{n\beta} \frac{g_{p_1 q}^{n\beta} g_{qp_2}^{n\beta}}{\hbar\Omega_{n\beta}} c^\dagger_{p_1\sigma} c_{p_2\sigma} 
\nonumber\\ 
-\, \sum_{\sigma\sigma'}\sum_{p_1p_2p_3p_4=1}^{2} \sum_{n\beta} \frac{g_{p_1p_2}^{n\beta} g_{p_3p_4}^{n\beta}}{\hbar\Omega_{n\beta}}    
c^\dagger_{p_1\sigma} c_{p_2\sigma} c^\dagger_{p_3\sigma'} c_{p_4\sigma'}.
\label{H_sys_2}
\end{eqnarray} 
Performing the normal ordering of the operators in the second line, we notice that the counter-term cancels out,
which results in the normal ordered renormalized Hamiltonian,
\begin{eqnarray}
&&\tilde H_{\rm sys} = H_{\rm sys} 
\nonumber\\ &&
-\, \sum_{\sigma\sigma'}\sum_{p_1p_2p_3p_4=1}^{2} \sum_{n\beta} \frac{g_{p_1p_2}^{n\beta} g_{p_3p_4}^{n\beta}}{\hbar\Omega_{n\beta}}    
c^\dagger_{p_1\sigma}  c^\dagger_{p_3\sigma'} c_{p_4\sigma'} c_{p_2\sigma}.
\nonumber\\ &&
\label{H_sys_normal}
\end{eqnarray}
Thus, we have demonstrated that the counter term, i.e. the second line of Eq. (\ref{Vint_normal_0}), restores the normal ordering of the fermionic operators in the renormalized Hamiltonian. 
This ordering is correct because all operators are normally ordered in the original Hamiltonian (\ref{H}).

Introducing the screened electron repulsion integrals
\begin{eqnarray}
\tilde h_{p_1p_2p_3p_4}  = h_{p_1p_2p_3p_4} - \sum_{n\beta}\frac{2g_{p_1p_2}^{n\beta} g_{p_3p_4}^{n\beta}}{\hbar\Omega_{n\beta}},
\label{h_screened} 
\end{eqnarray}
we can express the renormalized Hamiltonian  (\ref{H_sys_normal}) in the form similar to that of the original two orbital Hamiltonian (\ref{Hsys}),
\begin{eqnarray}
&& \tilde H_{\rm sys} = \sum_\sigma\sum_{p_1p_2=1}^{2} \tilde t_{p_1p_2} c^\dagger_{p_1\sigma} c_{p_2\sigma} 
\nonumber\\ &&
+ \frac{1}{2} \sum_{\sigma\sigma'} \sum_{p_1p_2p_3p_4=1}^{2} \tilde h_{p_1p_2p_3p_4} c^\dagger_{p_1\sigma} c^\dagger_{p_3\sigma'} c_{p_4\sigma'} c_{p_2\sigma}.
\label{H_static_2}
\end{eqnarray}
The renormalized parameters (\ref{param_static}) can be directly extracted from this Hamiltonian. For that purpose,
one should use the parameter definitions (\ref{U1U2},\ref{Delta_12},\ref{J_12},\ref{K_12}) and perform
the replacement $h_{p_1p_2p_3p_4} \to \tilde h_{p_1p_2p_3p_4}$ in them. 

The screened Coulomb integrals  (\ref{h_screened}) can be computed without diagonalizing the matrix $\hat M$ (\ref{M}),
it is sufficient to invert it, which is less computationally demanding. 
Indeed, substituting the coupling constants (\ref{g}) in Eq. (\ref{h_screened}) and performing simplifications in the 
sums, we obtain the screened integrals in the form
\begin{eqnarray}
&& \tilde h_{p_1p_2p_3p_4}  = h_{p_1p_2p_3p_4} -  \frac{4}{\hbar} \sum_{m_1\alpha_1,m_2\alpha_2}
 h_{p_1p_2 m_1\alpha_1} r_{m_1\alpha_1}  
\nonumber\\ && \times\,
\sqrt{\omega_{m_1\alpha_1}}
\big[ \hat M^{-1} \big]_{m_1\alpha_1,m_2\alpha_2} 
\sqrt{\omega_{m_2\alpha_2}} r_{m_2\alpha_2} h_{m_2\alpha_2 p_3p_4}.
\nonumber\\ 
\label{h_screened_2} 
\end{eqnarray}
The same transformations can be performed in the counter term appearing as the second term in the first line
of Eq. (\ref{H_sys_2}), after that it can be brought to the form given in Eq. (\ref{Vint_singlet}).
We also note that the screened integrals (\ref{h_screened_2}) match the corresponding expressions given in Ref. \cite{RPA}
if one puts $r_{m_1\alpha_1} = r_{m_2\alpha_2} = 1$.

\section{Benchmarking the static approximation of the system-bath model}

In Table \ref{table2} we list the excitation energies of the lowest triplet and the lowest singlet states
for a set of 20 molecules, which have been computed within the static screening approximation described in Sec. \ref{Sec_static}
and with molecular geometries provided by the QUEST database \cite{QUEST}.
We then compare these values with the reference values from QUEST. Detailed discussion
of this procedure is provided in Sec. \ref{results}. The last column of Table \ref{table2}
indicates the model of the environment used for the simulation. The symbol "sg" indicates
the molecule with the singlet ground state of the system Hamiltonian, 
and the symbol "tr" --- the molecules with the triplet ground state of $H_{\rm sys}$.

\end{document}